\documentclass[11pt,a4paper]{article}
\usepackage{jheppub,xcolor}
\pdfoutput=1
\usepackage[utf8]{inputenc}
\graphicspath{{./figs/}}

\usepackage{amssymb}
\usepackage{dcolumn}	
\usepackage{bm}			
\usepackage{bbm} 		
\usepackage{multirow}
\usepackage{slashed}
\usepackage{blkarray}
\usepackage{booktabs}
\usepackage{makecell}
\usepackage{tikz}
\usepackage[compat=1.1.0]{tikz-feynman}
\usepackage{longtable}
\usepackage{placeins}
\usepackage{listings}
\usepackage{amsmath}
\usepackage{xcolor}
\usepackage{pythonhighlight}
\usepackage{subcaption} 
\usepackage{rotating}
\usepackage{lscape}
\usepackage{setspace}
\usepackage{afterpage}

\def\gsim{\mathrel{\rlap{\lower4pt\hbox{\hskip1pt$\sim$}}
    \raise1pt\hbox{$>$}}}         
\def\lsim{\mathrel{\rlap{\lower4pt\hbox{\hskip1pt$\sim$}}
    \raise1pt\hbox{$<$}}}         

\definecolor{codegreen}{rgb}{0,0.6,0}
\definecolor{codegray}{rgb}{0.5,0.5,0.5}
\definecolor{codepurple}{rgb}{0.58,0,0.82}
\definecolor{backcolour}{rgb}{0.95,0.95,0.92}

\lstdefinestyle{mystyle}{
    backgroundcolor=\color{backcolour},   
    commentstyle=\color{codegreen},
    keywordstyle=\color{magenta},
    numberstyle=\tiny\color{codegray},
    stringstyle=\color{codepurple},
    basicstyle=\ttfamily\footnotesize,
    breakatwhitespace=false,         
    breaklines=true,                 
    captionpos=b,                    
    keepspaces=true,                 
    numbers=left,                    
    numbersep=5pt,                  
    showspaces=false,                
    showstringspaces=false,
    showtabs=false,                  
    tabsize=2
}
\definecolor{Red}{rgb}{1.,0.,0.}

\newcommand{\bc}{\begin{center}}
\newcommand{\ec}{\end{center}}
\newcommand{\ba}{\begin{array}}
\newcommand{\ea}{\end{array}}

\newcommand{\be}{\begin{equation}}
\newcommand{\ee}{\end{equation}}
\newcommand{\bea}{\begin{align}}
\newcommand{\eea}{\end{align}}

\def\smallfrac#1#2{\hbox{$\frac{#1}{#2}$}}
\def\half{\smallfrac{1}{2}}

\usepackage{tabularx}
\usepackage{multirow}

 \newcolumntype{C}[1]{>{\centering\arraybackslash}p{#1}}

\usepackage[normalem]{ulem}
\definecolor{ao}{rgb}{0.0, 0.5, 0.0}
\definecolor{mt}{RGB}{108, 210, 92}

\numberwithin{equation}{section}
\numberwithin{figure}{section}
\numberwithin{table}{section}

\title{A Determination of the Top Mass from a Global PDF Analysis}

\author[a]{Richard~D.~Ball,}
\author[a]{Jaco~ter~Hoeve,}
\author[a, b]{Roy~Stegeman}
\affiliation[a]{The Higgs Centre for Theoretical Physics, University of Edinburgh,\\[0.1cm]
JCMB, KB, Mayfield Rd, Edinburgh EH9 3FD, Scotland}
\affiliation[b]{Quantum Research Centre, Technology Innovation Institute, Abu Dhabi, UAE}

\emailAdd{r.d.ball@ed.ac.uk}
\emailAdd{jaco.ter.hoeve@ed.ac.uk}
\emailAdd{r.stegeman@ed.ac.uk}

\date{\today}

\abstract{
We present an indirect determination of the top-quark pole mass $m_t$ within a global analysis of parton distribution functions (PDFs),
based on the public NNPDF framework. We consider a wide range of  single- and double-differential top cross-section measurements, and compare them to theoretical predictions computed at NNLO QCD
accuracy with EW corrections, analysing their individual as well as combined impact on the joint
$(\alpha_s, m_t)$ parameter space, while accounting for PDF evolution up to
approximate ${\rm N^3LO}$ QCD accuracy with QED corrections. We account for missing higher order QCD uncertainties by default. Unique to our analysis are the inclusion of, first, toponium
contributions around the $t\bar{t}$ threshold, second, state-of-the-art
constraints on $\alpha_s$ from the lattice, and finally, a detailed sensitivity
study of the various ATLAS and CMS differential cross-section measurements at 8 and 13 TeV. We demonstrate
explicitly how a combined determination requires the refitting of the PDFs in order to correctly correlate uncertainties. We find $m_t = 172.81 \pm 0.26$ GeV at
approximate N$^3$LO QCD including NLO QED, EW and toponium corrections. 
}


\begin{document}
\begin{flushright}
\end{flushright}

\maketitle

\section{Introduction}
The top quark is the heaviest known particle in the Standard Model (SM) and is
copiously produced at the LHC. In Run II
alone, the ATLAS and CMS collaborations have collected an impressive number of
$t\bar{t}$ cross-section measurements, ranging from total cross sections
\cite{ATLAS:2022jbj, ATLAS:2014nxi, ATLAS:2020aln, CMS:2017zpm, CMS:2016yys,
CMS:2015yky}, to single~\cite{CMS:2018adi, ATLAS:2020ccu, ATLAS:2019hxz,
ATLAS:2016pal, ATLAS:2015lsn, CMS:2021vhb}, double~\cite{ATLAS:2020ccu,
ATLAS:2019hxz} and even triple~\cite{CMS:2019esx, CMS:2024ybg} differential measurements. Run III, operating at $\sqrt{s}=13.6$ TeV, has
already collected as much luminosity as all of Run II combined, and first
results are starting to become available~\cite{CMS:2023qyl, ATLAS:2023slx}. This
large data set enables an increasingly precise determination of the top quark
pole mass, $m_t$, with a current PDG average of $m_t = 172.4 \pm 0.7$ GeV
\cite{ParticleDataGroup:2024cfk}, see Ref.~\cite{Hoang:2020iah} for a review.

An accurate and precise determination of the top quark pole mass is relevant
for several reasons. It provides a key input parameter to many processes at
the LHC, and therefore tests the internal consistency of the SM
\cite{Baak:2014ora, deBlas:2021wap}. Moreover, being the heaviest SM particle,
it couples most strongly to the Higgs-boson and thus provides a promising avenue
to search for possible BSM effects~\cite{Kassabov:2023hbm, Celada:2024mcf}.
It also plays an important role, together with the strong coupling constant
$\alpha_s(m_Z)$, in determining whether the universe resides in a meta-stable or stable state
through the running of the quartic Higgs self-coupling~\cite{Andreassen:2017rzq,
Degrassi:2012ry, Bezrukov:2012sa, Bednyakov:2015sca}.

Following the discovery of the top quark at the Tevatron~\cite{CDF:1995wbb,
D0:1995jca}, the ATLAS and CMS collaborations have performed a series of
direct measurements of its mass~\cite{CMS:2018tye, CMS:2018fks, CMS:2019fak,
CMS:2021jnp, CMS:2023ebf, CMS:2022emx, ATLAS:2022jbw, ATLAS:2025bpp,
CMS:2022kqg}, of which $m_t = 172.95 \pm 0.53$~\cite{ATLAS:2025bpp} and
$m_t=171.77 \pm 0.37$ GeV~\cite{CMS:2023ebf} provide the most precise
direct determinations from a single channel.
The combination of ATLAS and CMS measurements up to run-I resulted in a
precision of $m_t=172.52 \pm 0.33$ GeV~\cite{ATLAS:2024dxp}. 
Looking beyond, High-Luminosity LHC (HL-LHC) projections currently predict a
sub-GeV precision of 200 MeV~\cite{ATLAS:2025eii}, which might be further
improved by future electron-positron colliders to a precision of
$\mathcal{O}(10)$ MeV~\cite{FCC:2025lpp}. 

In recent years, several groups outside the experimental collaborations have
performed determinations of $m_t$ from top production cross-sections, 
differing in terms of
experimental datasets, higher orders in the perturbative QCD expansion, and
methodologies~\cite{Alekhin:2017kpj, Garzelli:2020fmd, Cooper-Sarkar:2020twv,
Cridge:2023ztj, Garzelli:2023rvx, Holguin:2023bjf,Alekhin:2024bhs}. Given that
top cross-sections depend not only on $m_t$, but also on $\alpha_s$ and the parton distribution functions (PDFs), which themselves also 
depend strongly on $\alpha_s$, an accurate determination of $m_t$ from top cross-sections ideally requires a joint determination of all these quantities in the context of a global fit. This is necessary to capture all the correlations in the  
$(m_t,\alpha_s, {\rm PDF})$ parameter space: failing to take full account of such correlations can lead to biases~\cite{Forte:2020pyp,Forte:2025pvf}. 

The NNPDF collaboration recently performed a global determination of $\alpha_s$ from hadronic data, with a fixed value of $m_t$~\cite{Ball:2025xgq, Ball:2025xtj}. This determination was performed using the theory covariance method (TCM)~\cite{Ball:2018lag,Ball:2018twp,NNPDF:2019ubu,Ball:2021icz}, which 
enables accurate and efficient determination of discrete parameters such as 
$\alpha_s$ within the context of a global fit to PDFs, accounting for all correlations.
The aim of this paper is to likewise use the TCM to provide an accurate determination of the
top pole mass, $m_t$, while accounting for all correlations with $\alpha_s(m_Z)$ and the PDFs in the context of a global fit. We adopt state-of-the-art NNLO theoretical
predictions of $t\bar{t}$ production cross-sections, complemented with EW corrections, and benefit from the wide range of
experimental measurements included in NNPDF4.0~\cite{NNPDF:2021njg},
supplemented with several new $t\bar{t}$ measurements that have become available
since its release. We study in detail the sensitivity to $m_t$ to the various differential observables determined from the experimental datasets, and explicitly
demonstrate how their accurate combination can only be performed in the context of the global PDF fit. For the first
time, to the best of our knowledge, we also analyse how toponium corrections
\cite{ATLAS:2025mvr,CMS:2025kzt} affect the extraction of $m_t$.

The structure of this paper is as follows. First, Sect.~\ref{sec:meth} sets out the application of the Theory Covariance Method to the problem at hand.
Sect.~\ref{sec:data_theory} then describes the experimental datasets and
corresponding theoretical predictions that we use in our fit. Here we also
discuss how we model and account for (possible) toponium corrections.
Sect.~\ref{sec:results} presents the main results of our determination, first
validating our methodology in a controlled setup by means of a 
closure test, followed by a demonstration of the impact of
higher order perturbative corrections. We also comment on how our results
compare to previous results. Finally, we conclude and
summarise in Sect.~\ref{sec:summary}.

\section{Methodology} 
\label{sec:meth}

\subsection{The theory covariance method}
\label{subsec:tcm_method}

We start by reviewing the methodological framework that we adopt throughout this
work to extract $m_t$ alongside $\alpha_s(m_Z)$ from a global PDF fit. Our
methodology makes use of the theory covariance
matrix formalism~\cite{Ball:2018lag,Ball:2018twp,NNPDF:2019ubu}
used to account for theoretical uncertainties in PDF fits, and the Theory Covariance Method (TCM)~\cite{Ball:2021icz} which enables an analytical extraction of the underlying nuisance parameters. The TCM was originally developed to determine the correlations 
between theoretical uncertainties
entering PDF fits and predictions made using these PDFs, but it can
equally be used to optimise external physical parameters in the context of a PDF fit, as demonstrated explicitly in the case of $\alpha_s(m_Z)$
in Ref.~\cite{Ball:2025xgq}. The advantages of the TCM with respect to alternative methods are threefold: since the parameter extraction is analytic, as opposed to numerical, it does not interfere with the subtleties of the PDF extraction (in particular the cross-validation and stopping required because the PDFs are continuous functions); the covariance matrix of the parameters is determined without the need for a $\Delta \chi^2$ criterion, avoiding the tolerance ambiguities inherent in Hessian methods; the TCM is numerically accurate and efficient, requiring only a single PDF fit (as opposed to the multiple fits needed in the correlated replica method~\cite{Ball:2018iqk}), and as a result it makes it possible to extract any number of correlated external parameters simultaneously, without the need for exceptional computational resources. In this section we will show how the TCM can be applied to the special case of extracting two correlated physical parameters, the pole mass $m_t$ together with the strong coupling $\alpha_s(m_Z)$, in the context of a global PDF fit.

To this end, consider $n_{\rm dat}$ Gaussian distributed measurements $D=(D_1,
\dots, D_{n_{\rm dat}})^T$ (so $D$ is a column vector, while $D^T$ is a row vector), 
with covariance matrix $C$ in the space of the data, and the
corresponding theoretical predictions $T=(T_1, \dots, T_{n_{\rm dat}})^T$. Theory
uncertainties may be modelled by introducing univariate Gaussian nuisance
parameters $\lambda_a$ centered around zero, 
\begin{equation}
    P(\lambda) \propto \exp\left[-\half {\lambda_a^2}\right] \, ,
\end{equation}
whose effect is to smear out the vector of theory predictions $T$ by 
vectors $\beta_a$,
such that the conditional probability of $T$ is given by
\begin{equation}
P(T|D\lambda) \sim \exp\left[-\half(T + \lambda_a\beta_a-D)^T C^{-1}(T + \lambda_a\beta_a-D)\right] \, ,
\label{eq:prob_T_Dlambda}
\end{equation}
where the summation over the index $a$, labelling the nuisance parameters, is
left implicit. We assume all uncertainties to be
Gaussian distributed, which leaves the likelihood analytically tractable in the
nuisance parameters, and allows experimental uncertainties to be combined with estimates of purely theoretical uncertainties (such as missing higher
order terms, as demonstrated in Ref.~\cite{NNPDF:2024dpb}).
It is then straightforward to show that the introduction
of the nuisance parameters
$\lambda_a$ in Eq.~\eqref{eq:prob_T_Dlambda} is equivalent to adding a
contribution $\beta_a \beta_a^T$ to the original covariance matrix $C$.
Adopting a Bayesian framework, the probability $P(T|D)$ can be obtained by
marginalising $P(T|D\lambda)$ over $\lambda$,
\begin{equation}
    P(T|D) = \int d\lambda P(T|D\lambda)P(\lambda) \, ,
\end{equation}
which after completing the square in $\lambda$ leads to,
\be
P(T|D) \propto \exp\left[-\half(T-D)^T(C+S)^{-1}(T-D)\right] \, ,
\label{eq:prob_theory_data_2}
\ee
with $S=\beta_a\beta_a^T$. Therefore, theoretical uncertainties can be
accounted for simply by adding to the original
covariance matrix $C$ an additional contribution $S$. 

Moreover, we can also use Bayes' theorem to find the posterior distribution of
the $\lambda_a$:
\be
P(\lambda|TD) \propto
\exp\left[-\half(\lambda_{a}-\bar{\lambda}_a)Z_{ab}^{-1}(\lambda_{b}-\bar{\lambda}_b)\right]
\, ,
\label{eq:post_lambda}
\ee
where 
\begin{align} 
\label{eq:lambda_Z}
\bar{\lambda}_{a} &= \beta_a^T(C+S)^{-1}(D-T),\\
Z_{ab} &= \delta_{ab} - \beta_{a}^T(C+S)^{-1}\beta_b \, ,
\label{eq:Z_alphabeta}
\end{align}
denote the posterior mean and covariance of $\lambda_a$, respectively. Next we
consider some new theoretical prediction $\tilde{T}$ that has not been fitted to
$D$, although its theory uncertainties are fully correlated to those of $T$
(such that $\tilde{\lambda}_{a} = \lambda_a$), that is, 
\be
\tilde{T}(\lambda) = \tilde{T} + \lambda_a\tilde{\beta}_{a} \, .
\label{eq:T_tilde_gen}
\ee
Using Eq.~\eqref{eq:lambda_Z} we find that the posterior distribution on
$\lambda_a$ induces a shift $\delta \tilde{T}$ with respect to the prior
central value $\tilde{T}$,
\be
\delta \tilde{T} = \hat{S}(C+S)^{-1}(D-T) \, ,
\label{eq:delta_T_tilde}
\ee
where we have defined $\hat{S} \equiv \tilde{\beta}_a\beta_a^T$ as the
cross-covariance between the theoretical predictions entering the fit and the
new predictions $\tilde{T}$. Defining $\tilde{S} \equiv
\tilde{\beta}_a\tilde{\beta}^T_a$ as the prior theoretical covariance on
$\tilde{T}$, the posterior covariance on $\tilde{T}$ can be straightforwardly
obtained from Eq.~\eqref{eq:Z_alphabeta} as
\be 
\mathrm{Cov}[\tilde{T}] \equiv \tilde{S} - \hat{S}(C+S)^{-1}\hat{S}^T \, .
\label{eq:cov_T_tilde}
\ee
Note how the prior theoretical covariance $\tilde{S}$ is reduced by the second
term in Eq.~\eqref{eq:cov_T_tilde}, an example of statistical learning.

Let us now apply the above to the NNPDF methodology for determining the PDFs~\cite{NNPDF:2021njg,NNPDF:2021uiq}.
Rather than fitting $T$ directly to $D$, here we construct first $N_{\rm rep}$
data replicas $D^{(r)}$ drawn from a Gaussian distribution centered around the
experimental data $D$ with covariance $C + S$,
\begin{equation}
    D^{(r)} \sim \mathcal{N}(D, C + S) \, ,
\end{equation}
from which we then determine a corresponding theory replica $T^{(r)}$ by optimising
(through cross-validation) the following figure of merit as a function of the
PDFs $f$, 
\be
\chi^2_r[f] = (T^{(r)}[f]-D^{(r)})^T(C+S)^{-1}(T^{(r)}[f]-D^{(r)}) \, .
\label{eq:chi2_nnpdf}
\ee

While optimising Eq.~\eqref{eq:chi2_nnpdf} $m_t$ and $\alpha_s(m_Z)$ are kept
fixed at the prior central values 
$m_t^{0}$ and $\alpha_s^{0}$. We assume prior uncertainties $\Delta m_t$ and $\Delta \alpha_s$, and define $m_t^\pm = m_t^{0}\pm \Delta m_t$, $\alpha_s^\pm = \alpha_s^{0}\pm \Delta \alpha_s$. We then define four nuisance parameters as the allowable upward and downward variation of these two observables, as permitted by the PDF fit. 

We can now construct the vectors $\beta_a$, $a=1,\ldots.4$, and thus the three matrices $S$, $\hat{S}$ and $\tilde{S}$. We define $T^{(0)} \equiv \langle T^{(r)}\rangle$ as the central theory predictions obtained by taking the average over the replicas $r$. The vectors $\beta_a$ are constructed from four sets of shifts $\Delta T^\pm$ with respect to the central theory predictions $T^{(0)}$ induced by varying
$\alpha_s$ and $m_t$ around their prior central values (we choose four rather than two because the theory predictions are not necessarily linear in the external parameters):
\begin{align}
\Delta T_{m_t}^{\pm} &\equiv T^{(0)}\left(m_t^{\pm}\right) - T^{(0)}\left(m_t^{0}\right) \, ,\\
\Delta T_{\alpha_s}^{\pm} &\equiv T^{(0)}\left(\alpha_s^{\pm}\right) - T^{(0)}\left(\alpha_s^{0}\right) \, ,
\end{align}
and construct $\beta_a$ in terms of these,
\begin{equation}\renewcommand*{\arraycolsep}{0.1em}
    \beta_a = \frac{1}{\sqrt{2}}\begin{pmatrix}
\Delta T_{m_t}^+, & \Delta T_{m_t}^-, & \Delta T_{\alpha_s}^+, & \Delta T_{\alpha_s}^- \\
\end{pmatrix}	\, .
\label{eq:beta}
\end{equation}
Recalling that $S=\beta_a\beta_a^T$, we thus find for the prior covariance
of the theory predictions
\begin{equation}
S_{ij} = \half\left(\Delta T_{m_t, i}^+ \Delta T_{m_t, j}^+ + \Delta T_{m_t, i}^- \Delta T_{m_t, j}^- +\Delta T_{\alpha_s, i}^+ \Delta T_{\alpha_s, j}^+ + \Delta T_{\alpha_s, i}^- \Delta T_{\alpha_s, j}^- \right) \, ,
\end{equation}
where we have displayed the indices in data space explicitly, for clarity. 

Our two parameters to be predicted are $\tilde{T}(\lambda) = (m_t,\alpha_s)^T$, so we recast Eq.~\eqref{eq:T_tilde_gen} as
\be
\tilde{T} = \begin{pmatrix}
m_t^{0} + \lambda_{m_t^+}\Delta m_t - \lambda_{m_t^-}\Delta m_t\\
\alpha_s^{0} + \lambda_{\alpha_s^+}\Delta \alpha_s - \lambda_{\alpha_s^-}\Delta \alpha_s
\end{pmatrix}
\, ,\label{eq:replica_shifts}
\ee
where $\lambda_{m_t^\pm}$ and $\lambda_{\alpha_s^\pm}$ are the four nuisance parameters, to be determined replica by replica in the PDF fit optimising Eq.~\eqref{eq:chi2_nnpdf}. 
Then
\begin{equation}\renewcommand*{\arraycolsep}{0.1em}
    \tilde\beta_a = \frac{1}{\sqrt{2}}\begin{pmatrix}
\Delta m_t, & -\Delta m_t, & 0, & 0 \\
0, & 0, & \Delta \alpha_s, & -\Delta \alpha_s \\
\end{pmatrix}	\, ,
\label{eq:betatilde}
\end{equation}
and thus $\tilde{S} =\tilde{\beta}_a\tilde{\beta}^T_a$ while $\hat{S}= \tilde{\beta}_a\beta_a^T$,
\begin{equation}
    \tilde{S} = \begin{pmatrix}
\Delta m_t^2 & 0\\
0 & \Delta\alpha_s^2
\end{pmatrix} \, ,\qquad
    \hat{S}_i = \frac{1}{2}\begin{pmatrix}
(\Delta T^+_{m_t,i}-\Delta T^-_{m_t,i})\Delta m_t \\
(\Delta T^+_{\alpha_s,i}-\Delta T^-_{\alpha_s,i}) \Delta \alpha_s
\end{pmatrix} \,.
\end{equation}

Following similar steps that led to Eq.~\eqref{eq:delta_T_tilde}, the prior
central values $\tilde{T} = (m_t^0,\alpha_s^0)^T$ thus undergo the following net shifts
\be
\delta \tilde{T} = \hat{S}(C+S)^{-1}(D-T^{(0)}) \, .
\label{eq:delta_T_tilde_reps}
\ee
The associated
covariance  Eq.~\eqref{eq:cov_T_tilde} now receives an additional contribution arising 
from fluctuations of the replicas, corresponding to the PDF uncertainty: 
\begin{equation}
\mathrm{Cov}[\tilde{T}] = \tilde{S} - \hat{S}(C+S)^{-1}\hat{S}^T + \hat{S}(C+S)^{-1}X(C+S)^{-1}\hat{S}^T \, ,
\label{eq:cov_T_tilde_2}
\end{equation}
where $X$ is the covariance matrix of the optimal theory predictions determined by the fit,
\begin{equation}
    X_{ij} \equiv \langle (T_i^{(r)} - T_i^{(0)})(T_j^{(r)} - T_j^{(0)}\rangle \, ,
\end{equation}
the average again being over the replicas. The contribution from the PDF uncertainty is positive, because the necessity to use much of the data to determine the PDFs necessarily reduces the predictive power of the data in the determination of the external parameters.

Note that in the case
of a theory which always gives a perfect fit, 
that is $T=D$ for all replicas, we would have $X=C+S$ and
Eq.~\eqref{eq:cov_T_tilde_2} reduces to the prior uncertainty
$\tilde{S}$. This is because in that case all information contained in the data is
absorbed completely by the theory, leaving no room to inform the nuisance
parameters so that the posterior can only coincide with the prior. In practice,
a theory like the Standard Model is highly constraining, and thus cannot always fit data fluctuations exactly, leaving sufficient wiggle room to update the external parameters. In a sense this is what we mean by a predictive theory, as opposed to a purely phenomenological one.

Regarding the prior uncertainties on $\alpha_s(m_Z)$ and $m_t$, we choose sensible values to give a distribution wide enough that the 
final results are independent of the prior, but not so wide as to make the  
dependence of the theory predictions nonlinear.  In practice, we typically take $\Delta m_t = 2.5$ GeV, $\Delta \alpha_s = 0.002$, motivated by the fact that this comfortably encloses 
uncertainties in the PDG average~\cite{ParticleDataGroup:2024cfk}.

\subsection{Closure test methodology}
\label{subsec:closure-test}
Before applying the TCM from Sect.~\ref{subsec:tcm_method} to real data, one
must first verify whether it is free from any methodological bias by means of a
closure test~\cite{NNPDF:2014otw,NNPDF:2021njg,DelDebbio:2021whr,Harland-Lang:2024kvt,Barontini:2025lnl}. In this approach, synthetic datasets are first
 generated from a known underlying law $T^*\equiv T(\alpha_s^*,m_t^*,f^*)$ with chosen values $\alpha_s^*$ and $m_t^*$, and a fixed PDF $f^*$, and with a covariance $C+S$. The TCM methodology is then
applied on each of the synthetic data sets, producing a series of best-fit
$\alpha_s$ and $m_t$ values that should be statistically compatible with the
true underlying values if the closure test is successful.

In particular, one generates $N_{L_1}$ replicas $D_{L_1}^{(k)}$ drawn from a
Gaussian distribution centred around $T^*$ with covariance $C+S$, 
\begin{equation}
    D^{(k)}_{L_1} \sim \mathcal{N}\left(T^*, C+S\right)\,,\quad k=1, \dots ,N_{L_1}\, , 
    \label{eq:L1-data}
\end{equation}
where the $k^{\rm th}$ replica $D_{L_1}^{(k)}$ corresponds to the experimental
central value one would obtain if one were to imagine performing repeated ``runs
of the universe", each time observing a different sample of central values. In the standard NNPDF methodology the uncertainties are propagated by bootrapping the experimental data, for this reason we introduce a second level of noise, denoted $L_2$,
and draw $N_{L_2}$ replicas, this time from each of the $L_1$-replicas,
\begin{equation}
    D_{L_2}^{(k, r)} \sim \mathcal{N}\left(D_{L_1}^{(k)}, C+S\right)\,,\quad r=1, \dots ,N_{L_2}\, ,
    \label{eq:L2-data}
\end{equation}
Then, for each $k$, we perform a fit following the methodology from
Sect.~\ref{sec:meth} to end up with pairs $\tilde{T}^{(k)}\equiv(
m_t^{(k)}, \alpha_s^{(k)})$, each with corresponding covariance ${\rm Cov}[
m_t^{(k)}, \alpha_s^{(k)}]$. The weighted average of the collection of pairs is then given by
\begin{equation}
    \langle (m_t, \alpha_s)\rangle = \left(\sum_{k=1}^{N_{L_1}}{\rm Cov}^{-1}[m_t^{(k)}, \alpha_s^{(k)}]\right)^{-1}\left(\sum_{k=1}^{N_{L_1}}{\rm Cov}^{-1}[m_t^{(k)}, \alpha_s^{(k)}]\cdot \tilde{T}^{(k)}\right) \, ,
    \label{eq:mean_closure}
\end{equation}
with covariance,
\begin{equation}
    {\rm Cov}[\langle(m_t, \alpha_s)\rangle] \equiv \left(\sum_k {\rm Cov}^{-1}[m_t^{(k)}, \alpha_s^{(k)}]\right)^{-1} \, .
    \label{eq:cov_closure}
\end{equation}
For a successful closure test, Eqs.~\eqref{eq:mean_closure} and
\eqref{eq:cov_closure} must be statistically compatible with the true underlying
values $\alpha_s^*$ and $m_t^*$. The results of such a closure test will be presented in
Sect.~\ref{subsec:closure-test-results}.
\section{Experimental data and theoretical predictions}
\label{sec:data_theory}
In this Section, we describe the experimental data and corresponding theoretical
predictions that enter our analysis. We take the NNPDF4.0 data set as our
baseline~\cite{NNPDF:2021njg}, complemented wherever possible with new
$t\bar{t}$ production data sets that have become available since its release. 
Single top data are not included in this study. In the
following, we therefore focus exclusively on $t\bar{t}$ cross-section data 
and refer to Ref.~\cite{NNPDF:2021njg} for a complete overview regarding all the other 
processes included in the global extraction of $m_t$.

\subsection{Experimental measurements}

Given the wide variety of the top measurements from LHC, it is important to 
distinguish between what we mean by an
experiment, a data set and an observable. We refer to ATLAS and CMS as
experiments, and each provides data sets at different centre of mass energies (8 or 13 TeV)
and different decay channels (dilepton, lepton + jet, or hadronic). 
Within each data set we can further distinguish between
cross-section measurements differential in different kinematic observables, such as 
the top invariant mass $m_{t\bar{t}}$, the
top transverse momentum $p_T^t$, top quark rapidities $y_t$ or $y_{t\bar{t}}$,
or combinations thereof in the case of double
differential distributions. We refer to a measurement collectively as a
particular combination of an experiment (ATLAS/CMS), data set ($\sqrt{s}$ and
channel) and observable (differential in $m_{t\bar{t}}$, $p_T^t$, etc.). In general,
measurements at different experiments and at different energies and/or channels
can be assumed to be independent, but measurements of different observables within
the same experiment and data set are not. The latter may thus only enter a
global PDF fit simultaneously when their cross-correlation is (publicly)
available.

Table \ref{tab:datasets} provides an overview of the $t\bar{t}$ measurements
that we consider, where, for each measurement, we indicate the available
observables, the number of data points, $n_{\rm dat}$, the total integrated
luminosity, $\mathcal{L}_{\rm int}$, and its corresponding reference. We use only differential un-normalized observables in this study, and don't include the total cross-section measurements since these are not independent.
\begin{table}[htbp]
  \centering
  \scriptsize
  \renewcommand{\arraystretch}{1.43}
  \begin{tabularx}{\textwidth}{Xlccc}
\toprule
\bf{Experiment and data set}
& \bf{Observable}
& $n_{\rm dat}$
& $\mathcal{L}_{\rm int}$ $[{\rm fb}^{-1}]$
& \bf{Ref.}
\\
\midrule
\multirow{3}{*}{ATLAS 13~TeV $t\bar{t}$ all hadr.} & $d\sigma/dm_{t\bar{t}}$ &  9 &\multirow{3}{*}{36.1}  & \multirow{3}{*}{\cite{ATLAS:2020ccu}}\\
&$d\sigma/d|y_{t\bar{t}}|$ & 12  & \\
&$d^2\sigma/dm_{t\bar{t}}\ d|y_{t\bar{t}}|$ & 11  & \\
\midrule
\multirow{5}{*}{ATLAS 13~TeV $t\bar{t}$ $\ell+j$} & $d\sigma/dm_{t\bar{t}}$ & 9  &\multirow{5}{*}{36.0} & \multirow{5}{*}{\cite{ATLAS:2019hxz}} \\
&  $d\sigma/dp_T^t$ & 8  & \\
&  $d\sigma/ d|y_{t}|$ & 5  & \\
&  $d\sigma/d|y_{t\bar{t}}|$ & 7   & \\
& $d^2\sigma/dm_{t\bar{t}}\ dp_T^t$ & 15   & \\
\midrule
\multirow{2}{*}{ATLAS 8~TeV $t\bar{t}$ $2\ell$} &
 $d\sigma/dm_{t\bar{t}}$ & 6  & \multirow{2}{*}{20.2} & \multirow{2}{*}{\cite{ATLAS:2016pal}} \\
& $d\sigma/dy_{t\bar{t}}$ &  5  & \\
\midrule
\multirow{4}{*}{ATLAS 8~TeV $t\bar{t}$ $\ell+j$ } &
$d\sigma/dm_{t\bar{t}}$ & 7   & \multirow{4}{*}{20.3} & \multirow{4}{*}{\cite{ATLAS:2015lsn}} \\
& $d\sigma/dp_T^t$ & 8  & \\
& $d\sigma/dy_t$ & 5   & \\
& $d\sigma/dy_{t\bar{t}}$ & 5  & \\
\midrule
\multirow{4}{*}{CMS 13~TeV $t\bar{t}$ $2\ell$ 138 ${\rm fb}^{-1}$}  & $d\sigma/dm_{t\bar{t}}$ &  7   &\multirow{4}{*}{138.0}  & \multirow{4}{*}{\cite{CMS:2024ybg}}\\
&$d^2\sigma/dm_{t\bar{t}}\ d|y_{t\bar{t}}|$ & 16 & \\
&$d\sigma/dp_T^t$ & 7   & \\
&$d\sigma/d|y_{t}|$ &  10   & \\
\midrule
\multirow{4}{*}{CMS 13~TeV $t\bar{t}$ $2\ell$} & $d\sigma/dm_{t\bar{t}}$ & 7  &\multirow{4}{*}{35.9} & \multirow{4}{*}{\cite{CMS:2018adi}} \\
&  $d\sigma/dp_T^t$ & 6  & \\
&  $d\sigma/ d|y_{t}|$ & 10   & \\
&  $d\sigma/d|y_{t\bar{t}}|$ & 10   & \\
\midrule
\multirow{5}{*}{CMS 13~TeV $t\bar{t}$ $\ell+j$ } &
$d\sigma/dm_{t\bar{t}}$ & 15  & \multirow{5}{*}{137.0} & \multirow{5}{*}{\cite{CMS:2021vhb}} \\
&$d^2\sigma/dm_{t\bar{t}}\ d|y_{t\bar{t}}|$ & 35  &   & \\
& $d\sigma/dp_T^t$ & 16   & \\
& $d\sigma/dy_t$ & 11  & \\
& $d\sigma/dy_{t\bar{t}}$ & 10   & \\
\bottomrule
\end{tabularx}

  \vspace{0.3cm}
  \caption{Overview of the $t\bar{t}$ measurements considered in the current
  analysis. For each measurement, we indicate the available observables, the
  number of data points, $n_{\rm dat}$, the integrated luminosity,
  $\mathcal{L}_{\rm int}$ and the corresponding reference.}
  \label{tab:datasets}
\end{table}

Specifically, regarding measurements from ATLAS, we include the fully hadronic channel at 13
TeV with a luminosity of $36.1$ ${\rm fb}^{-1}$~\cite{ATLAS:2020ccu}, single
differential in $m_{t\bar{t}}$, the absolute rapidity $|y_{t\bar{t}}|$, and the
double-differentially in $(m_{t\bar{t}}, |y_{t\bar{t}}|)$. In the $\ell + j$
channel, we include the 13 TeV measurement at $36.0$ ${\rm fb}^{-1}$ from
Ref.~\cite{ATLAS:2019hxz}, single differential in $m_{t\bar{t}}$, $p_T^t$,
$|y_t|$, $|y_{t\bar{t}}|$ and double differential in $(m_{t\bar{t}}, p_T^t)$. At
8 TeV, we include the dilepton measurement from Ref.~\cite{ATLAS:2016pal} at an
integrated luminosity of 20.2 ${\rm fb}^{-1}$, single differential in
$m_{t\bar{t}}$ and $y_{t\bar{t}}$. In the $\ell + j$ channel, we include the
measurement from Ref.~\cite{ATLAS:2015lsn} single differential in $m_{t\bar{t}},
p_T^t$ and the rapidities $y_t$ and $y_{t\bar{t}}$.

In the case of CMS, we consider the dilepton measurement based on the full
run-II luminosity from Ref.~\cite{CMS:2024ybg}, single differential in
$m_{t\bar{t}}$, $p_T^t$, and the (absolute) rapidities $y_t$ and $y_{t\bar{t}}$,
and double differential in the pair $(m_{t\bar{t}}, |y_{t\bar{t}}|)$. We also
consider the corresponding lower luminosity measurement at 35.9 ${\rm fb}^{-1}$
from Ref.~\cite{CMS:2018adi} for those observables that are not available from
the full Run-II equivalent. In the $\ell + j$ channel, we include the run-II
legacy measurement from Ref.~\cite{CMS:2021vhb} single differential in
$m_{t\bar{t}}$, $p_T^t$, $y_t$ and $y_{t\bar{t}}$, and double differential in
the pair $(m_{t\bar{t}}, |y_{t\bar{t}}|)$. At 8 TeV CMS only provide normalized differential distributions, which we thus do not include.

\subsection{Theory settings}
\label{subsec:theory_settings}
For each of the measurements tabulated in Table \ref{tab:datasets}, we compute
theoretical predictions at NNLO QCD accuracy using \verb|MATRIX|
\cite{Devoto:2025cuf, Grazzini:2017mhc} interfaced to the {\sc\small PineAPPL}
fast grid interface~\cite{Carrazza:2020gss, christopher_schwan_2025_15635174}.
All measurements are therefore analysed at parton level.
Following the methodology outlined in Sect.~\ref{sec:meth}, we produce three
sets of theoretical predictions, one for each value of the pole mass $m_t$ in the
on-shell scheme; the prior central value $m_t^{0}=172.5\,{\rm GeV}$, the up
variation, $m_t^{+} = 175.0 \, \rm{GeV}$, and the down variation at $m_t^{-} =
170.0 \, \rm{GeV}$. This constitutes our prior, as defined in
Eq.~\eqref{eq:beta}, and its variations are motivated by the fact that the PDG
average, $m_t = 172.4 \pm 0.7\,{\rm GeV}$, falls comfortably inside it
\cite{ParticleDataGroup:2024cfk}. Regarding $\alpha_s$ variations, we produce
predictions for our entire data set, so including all non $t\bar{t}$ data, at 
\be
 \alpha_s = \{0.116, 0.117, \dots, 0.124, 0.125\} \, ,
 \ee
 giving us flexibility to redefine our $\alpha_s$ prior as needed.  We have benchmarked our
 predictions at NNLO against the database \verb|hightea|~\cite{Czakon:2023hls},
 as well as \textsc{MadGraph5\_aMC@NLO}~\cite{Alwall:2014hca} in the case of NLO QCD,
 finding perfect agreement within uncertainties in both cases. The
 renormalisation and factorisation scales are set to $\mu_R=\mu_F = H_T/4$, with
 $H_T$ the transverse energy of the $t\bar{t}$ event, defined as,
\begin{equation}
    H_T \equiv \sqrt{p_{T, t}^2 + \left(m_t^{0}\right)^2} +  \sqrt{p_{T, \bar{t}}^2 + \left(m_t^{0}\right)^2} \, .
    \label{eq:dyn_scale}
\end{equation}
Note that Eq.~\eqref{eq:dyn_scale} is always evaluated at $m_t=m_t^{0}$, meaning that
we exclusively vary the top mass dependence in the hard scattering, while
leaving $\mu_R$ and $\mu_F$ fixed to Eq.~\eqref{eq:dyn_scale}. In this way, we
probe the true parametric dependence on the top mass, separating it from the scale 
dependence, which we use to estimate missing higher order uncertainties (MHOUs).

\begin{figure}[ht!]
  \begin{subfigure}[b]{0.49\textwidth}
      \centering
      \includegraphics[width=\textwidth]{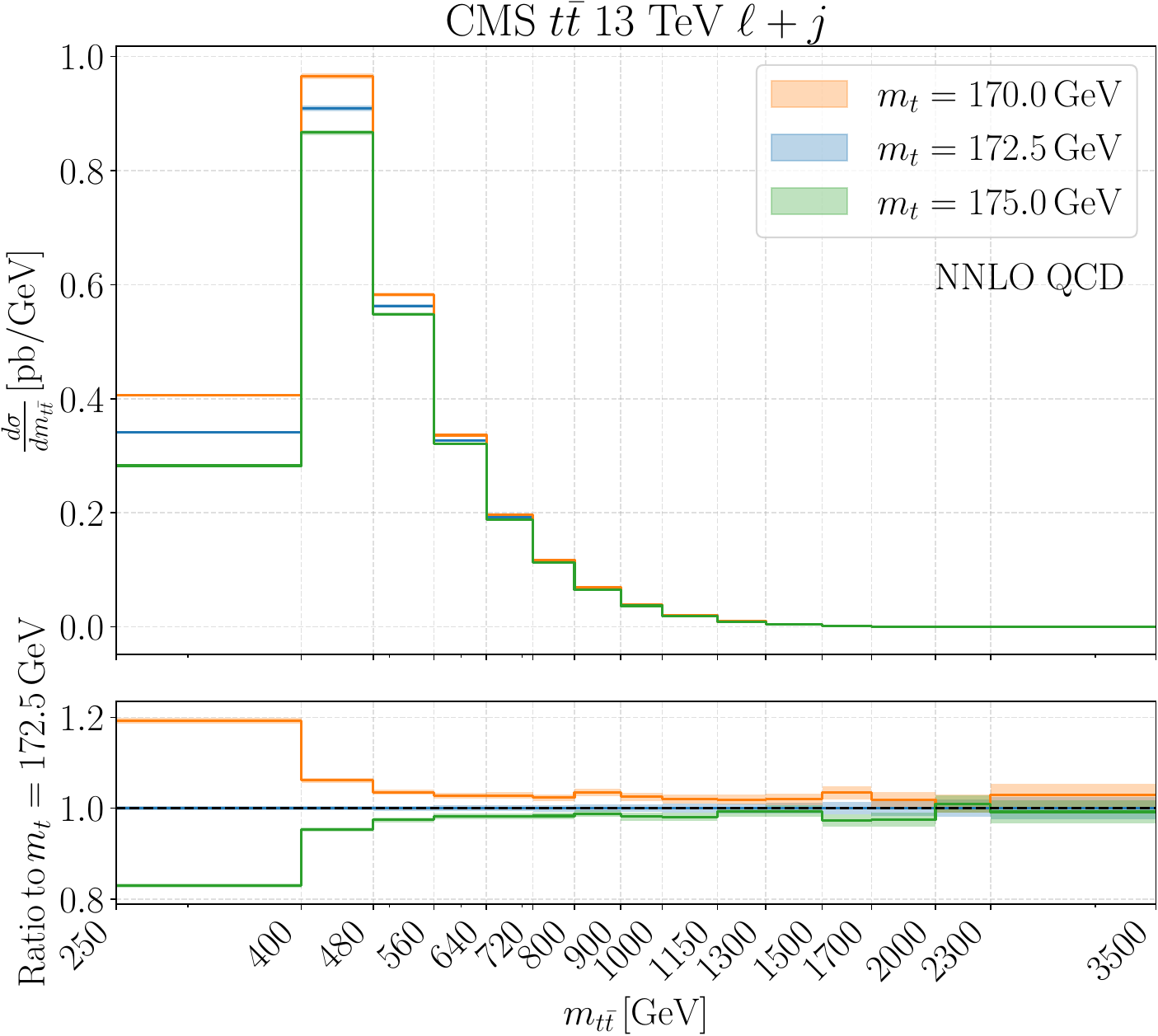}
  \end{subfigure}
  \hfill
  \begin{subfigure}[b]{0.49\textwidth}
      \centering
      \includegraphics[width=\textwidth]{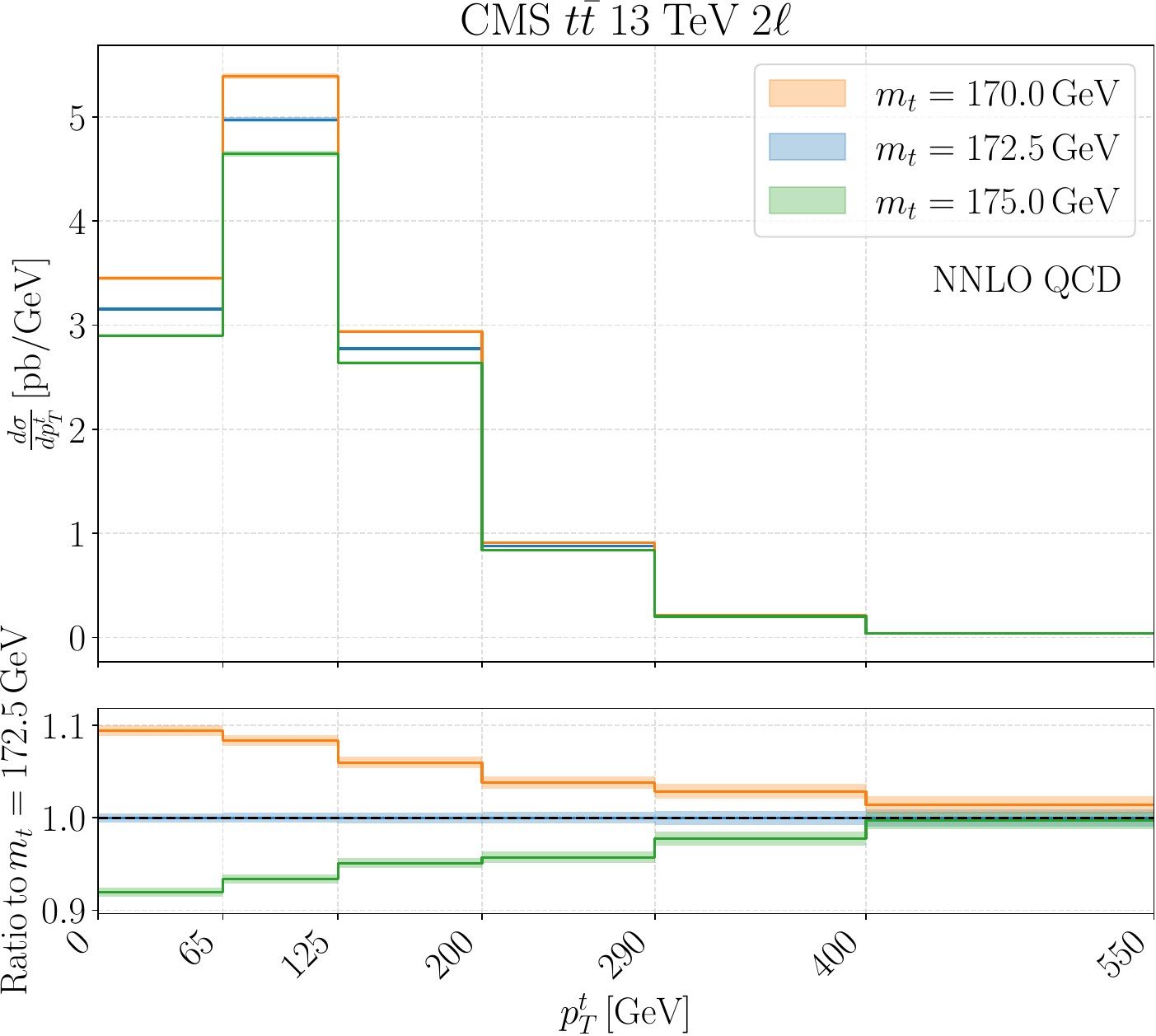}
  \end{subfigure}

  \vskip\baselineskip
  \begin{subfigure}[b]{0.49\textwidth}
      \centering
      \includegraphics[width=\textwidth]{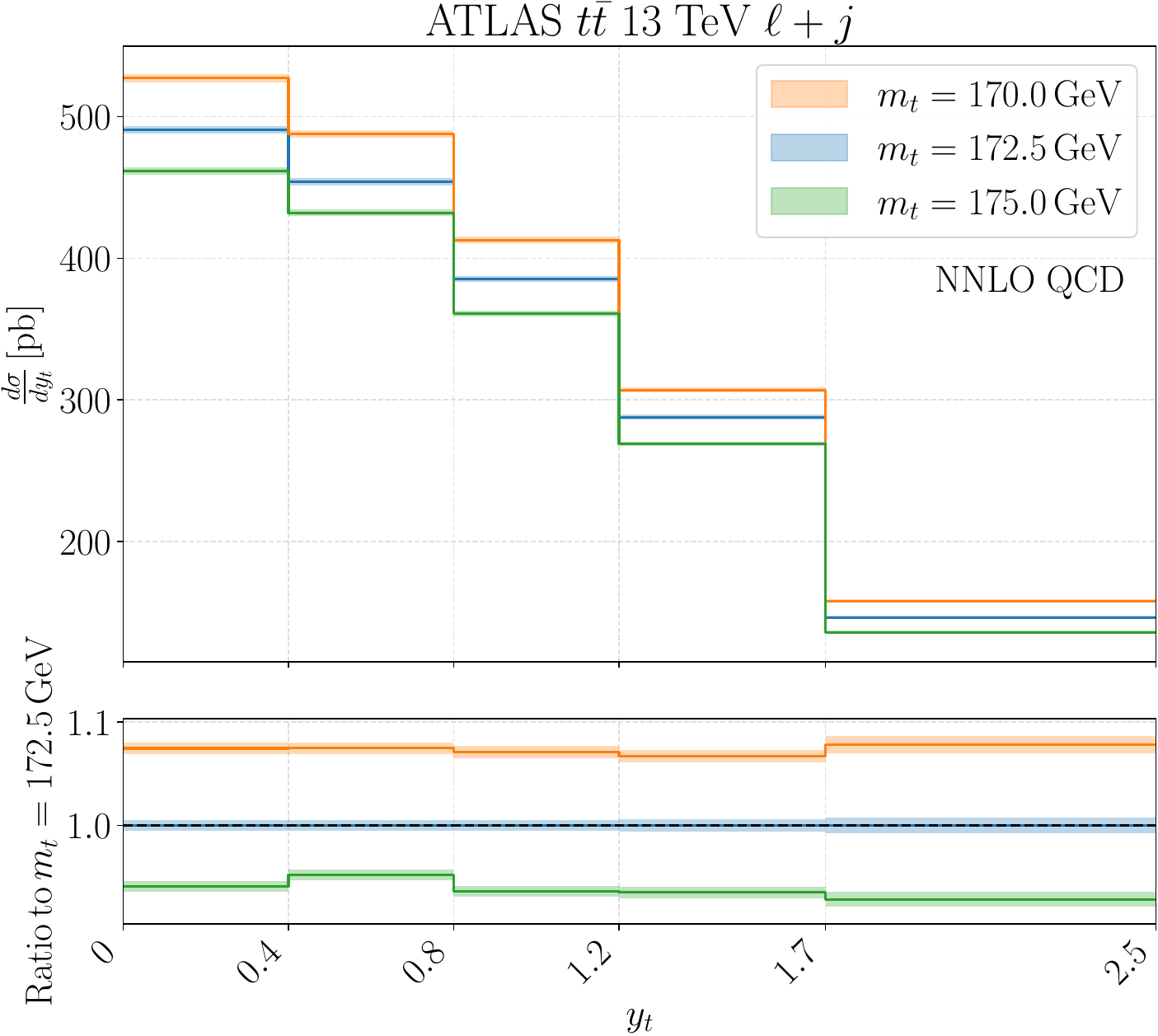}
  \end{subfigure}
  \hfill
  \begin{subfigure}[b]{0.49\textwidth}
      \centering
      \includegraphics[width=\textwidth]{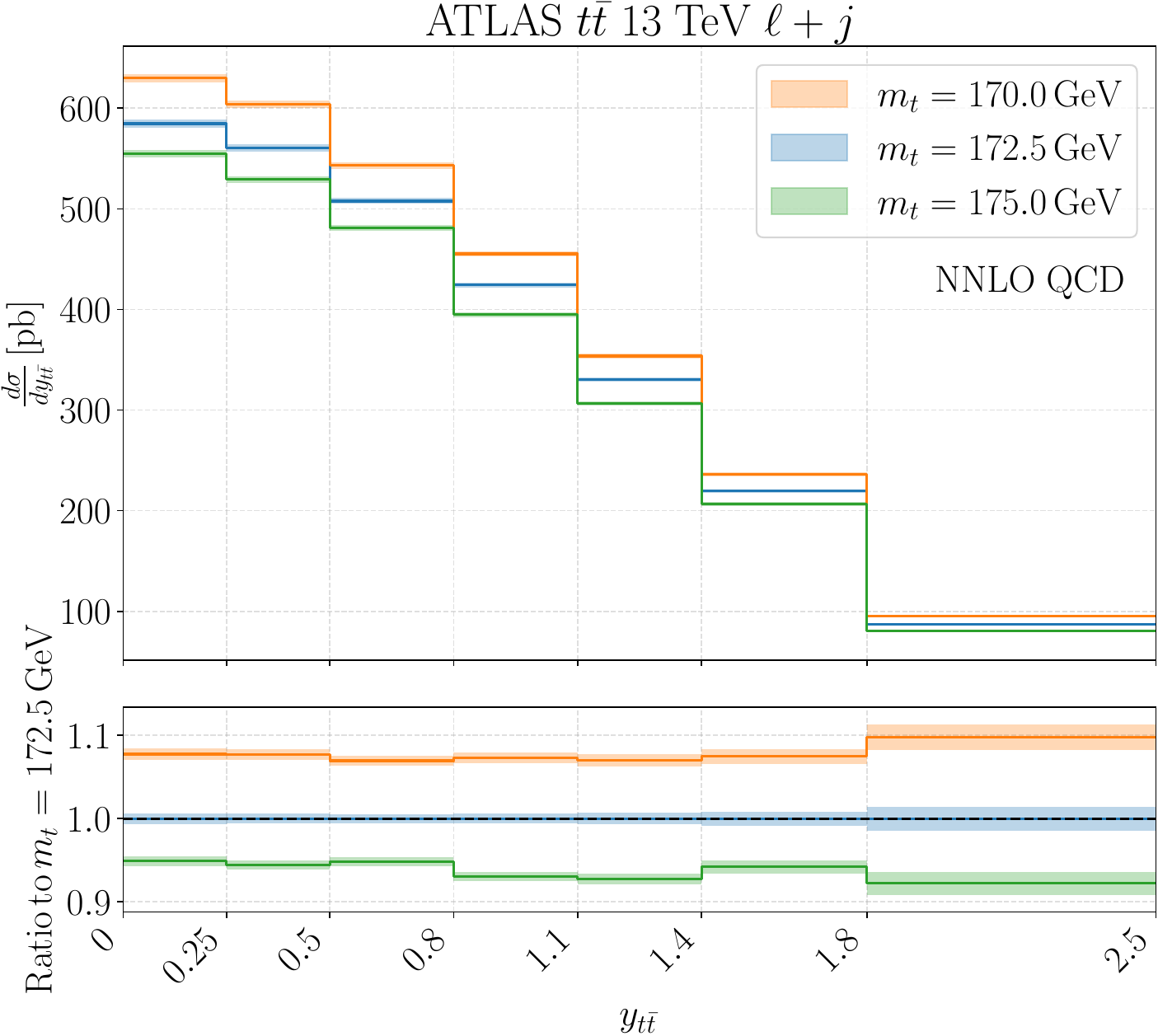}
  \end{subfigure}

  \caption{Representative differential distributions of $t\bar{t}$ at 13 TeV by ATLAS and CMS at the top-quark pole mass $m_t = \{170.0, 172.5,
  175.0\}$ GeV at NNLO QCD. From top left to bottom right, we show distributions
  differential in the invariant mass $m_{t\bar{t}}$, the transverse momentum
  $p_T^t$, top-quark rapidity $y_t$ and the rapidity $y_{t\bar{t}}$ of the
  top-quark pair. The shaded bands represent the PDF uncertainties obtained from
  NNPDF4.0~\cite{NNPDF:2021njg}}
  \label{fig:kinematic_dist}
\end{figure}

While we have made sure that Monte Carlo (MC) uncertainties on the total
integrated cross-sections are below $0.2\%$, which translates into sub-percent
precision across all differential bins, the relative uncertainty expressed with
respect to the up and down variations $\Delta T_{m_t, i}^{\pm}$ can be
considerably larger, reaching e.g. $\mathcal{O}(10\%)$ in high invariant mass
bins. To remedy this, we account for MC fluctuations by adding the following
(diagonal) contribution to the fit covariance matrix,
\begin{equation}
    C_{ij}^{\rm (MC)} = \frac{1}{2}\left((\delta T_{i, m_t}^+)^2 + (\delta T_{i, m_t}^-)^2\right)\delta_{ij} \, ,
    \label{eq:mc_covmat}
\end{equation}
where $\delta T_{i, m_t}^\pm$ denotes the MC uncertainty on up and downward
varied $t\bar{t}$ theory predictions.

It is interesting to visualise how each of the kinematic distributions from
Table \ref{tab:datasets} are affected by varying $m_t$ around its nominal
value. Fig.~\ref{fig:kinematic_dist} displays a representative subset of
$t\bar{t}$ differential distributions at 13 TeV for $m_t = \{170.0,
172.5, 175.0\}$ GeV. From top left to bottom right, we show distributions in the
invariant mass $m_{t\bar{t}}$, the transverse momentum $p_T^t$ of the top-quark
and the rapidities $y_t$ and $y_{t\bar{t}}$. In each case, the shaded colour
band represents the PDF uncertainties from the
\verb|NNPDF40_nnlo_as_01180| PDF set~\cite{NNPDF:2021njg}. Several observations
can be made by inspecting Fig.~\ref{fig:kinematic_dist}. First, the
$m_{t\bar{t}}$ distribution (top left panel) is clearly the most sensitive to
$m_t$, specifically just above the $t\bar{t}$ threshold where variations of 
up to $20\%$ in both directions are observed as indicated in the ratio-plot.  At higher invariant masses the shifts drop to
below $5\%$ making this region of phase space less sensitive. Similar
effects are observed in the $p_T^t$ distribution (top right panel), where $m_t$
variations induce shifts between typically $5\%$ and $10\%$ just above threshold, with only
very small shifts at high $p_T$. A rather different picture emerges in
the rapidity distributions shown in the lower panels. Here, no shape
effects are observed, i.e. the cross-section shows roughly comparable shifts across all
differential bins. We therefore expect the $m_{t\bar{t}}$ and $p_T^t$
distributions to give the most stringent constraints on $m_t$, with the 
rapidity distributions being rather less sensitive. 

In the following, we always include MHOUs on the theory predictions following
Refs.~\cite{NNPDF:2024dpb,NNPDF:2024nan} in the 7-point scheme. We adopt the
EKO~\cite{Candido:2022tld, candido_2025_15655642} evolution code for PDF
evolution, where QED corrections are included following
Ref.~\cite{NNPDF:2024djq}, and aN$^3$LO QCD corrections are implemented
according to Ref.~\cite{NNPDF:2024nan} complemented with the recent heavy quark
matching functions of Ref.~\cite{Ablinger:2024xtt} and splitting functions of
Ref.~\cite{Falcioni:2024qpd}.

\begin{figure}[t]
    \centering
    \includegraphics[width=0.9\linewidth]{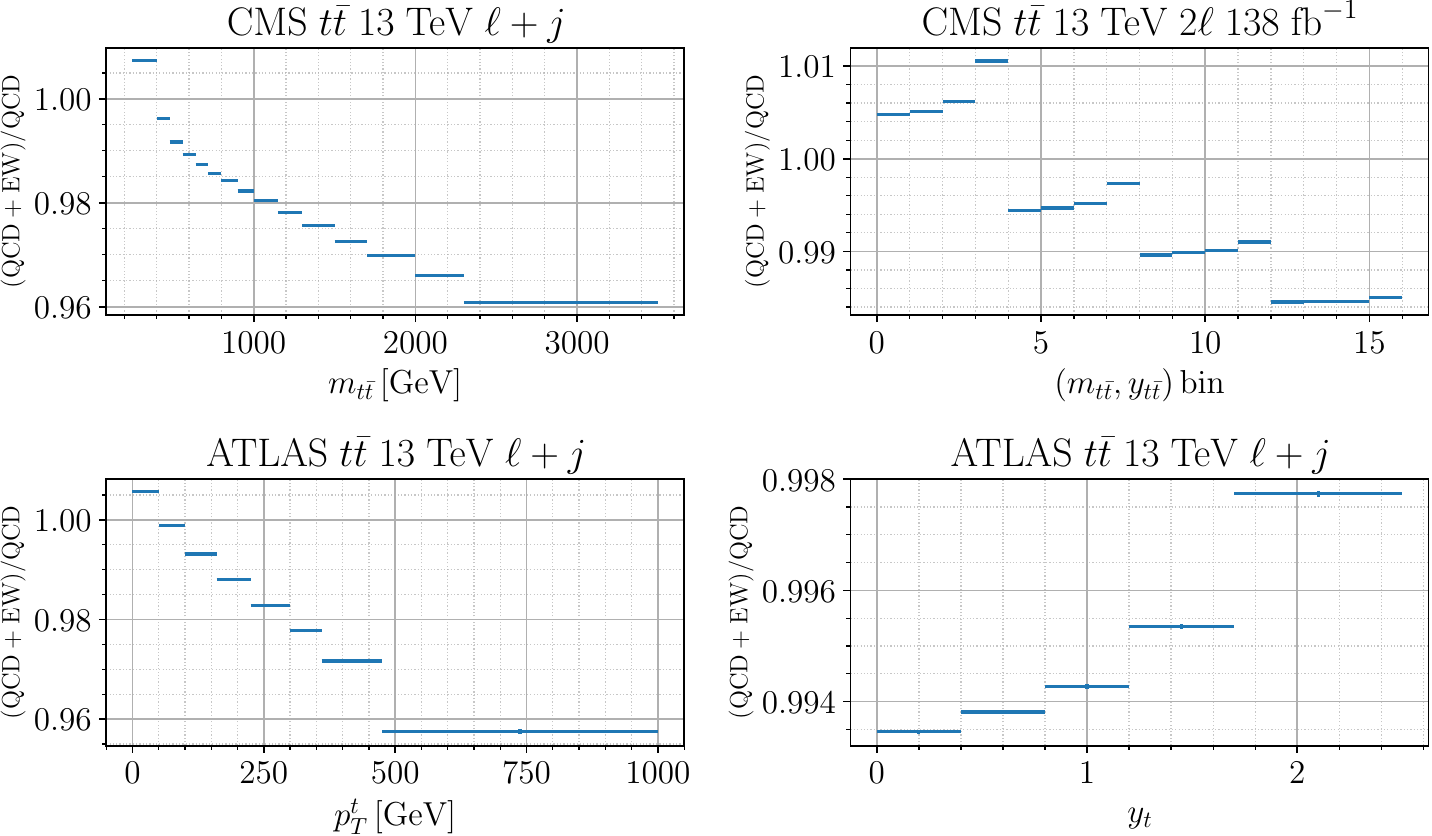}
    \caption{Relative EW corrections when considered on top of the NNLO QCD
    theory for a representative subset of distributions considered in this work. 
    From top left to bottom right, we display the impact in the $m_{t\bar{t}}$
    distribution, the double differential
    distribution in $(m_{t\bar{t}}, y_{t\bar{t}})$, the $p_T^t$ distribution,
    and the $y_t$ distribution
    \label{fig:ewk-factors}
    }
\end{figure}

For all distributions considered in this work, we compute electroweak
corrections using \textsc{MadGraph5\_aMC@NLO}~\cite{Alwall:2014hca}.
Our conventions
are similar to those adopted in Ref.~\cite{Carrazza:2020gss} and follow the
additive approach as defined in Ref.~\cite{Czakon:2017wor} for combining NNLO
QCD and EW corrections. In
particular, our NNLO QCD + EW computation includes the $\mathcal{O}(\alpha_s^2)$,
$\mathcal{O}(\alpha_s^3)$ and $\mathcal{O}(\alpha_s^4)$ QCD contributions, and
the LO $\mathcal{O}(\alpha_s\alpha)$ and NLO $\mathcal{O}(\alpha_s^2\alpha)$
contributions. We do not include the NLO contributions of order
$\mathcal{O}(\alpha_s\alpha^2)$, nor the purely EW corrections of order
$\mathcal{O}(\alpha^2)$ and $\mathcal{O}(\alpha^3)$. We adopt the same dynamical
scale choice as Eq.~\eqref{eq:dyn_scale} and use the
\verb|NNPDF40_nlo_as_01180_qed| PDF set~\cite{NNPDF:2021njg} to include photon
initiated channels. Fig.~\ref{fig:ewk-factors} displays the impact of the EW corrections expressed relative to the NNLO
QCD theory for a representative subset of the $m_{t\bar{t}}$, $p_T^t$ and rapidity
distributions from Table \ref{tab:datasets}. Similarly to what was argued in 
Ref.~\cite{Czakon:2017wor}, we note how the relative size of the EW corrections depends strongly on the specific kinematic distribution
considered, with the largest impact of around -4\% observed in the case of the
$m_{t\bar{t}}$ and $p_T^t$ distributions, while the rapidity distributions only
undergo rescalings at the level of a few per mille.

\subsection{Toponium corrections} 
\label{subsec:toponium_corrections}
An intriguing property in $t\bar{t}$ production that has attracted significant
interest recently is the formation of toponium just below the top-pair
production threshold. This colour-singlet $t\bar{t}$ quasi-bound state was
observed independently by the ATLAS and CMS collaborations with an observed cross-section of $9.3^{+1.4}_{-1.3}$ pb and
$8.8^{+1.2}_{-1.4}$ pb, respectively~\cite{ATLAS:2025mvr,
CMS:2025kzt}. Its effect shows up most notably in the $t\bar{t}$
invariant mass spectrum as a smeared out bump before threshold, where it enhances the
$t\bar{t}$ production cross section by around 6.43 pb at $\sqrt{s} = 13$ TeV
according to theoretical estimates
\cite{Fuks:2021xje}. This raises the question whether it may have significant
impact on the top-quark mass determination, especially given how the latter is
predominantly sensitive to the threshold region as observed in
Fig.~\ref{fig:kinematic_dist}. Even though its relative contribution to the
overall cross-section is below 1\%~\cite{Fuks:2021xje}, we remark that it can
lead to contributions as high as 6\% when considered in isolated bins around
threshold, which is clearly significant.

To this end, we account for toponium effects by modelling its contribution
through non-relativistic QCD (NRQCD) following the steps outlined in
Ref.~\cite{Fuks:2024yjj}, and as first discussed in Refs.~\cite{Fadin:1987wz, Fadin:1990wx}. In particular, we adopt \textsc{MadGraph5\_aMC@NLO}
\cite{Alwall:2014hca} and generate 100K LO events in the gluon initiated
channel,
\begin{equation}
    g g \rightarrow t\bar{t} \rightarrow b\ell^+\nu_\ell\bar{b}\ell'^-\bar{\nu}'_\ell \, ,
    \label{eq:toponium_mg5}
\end{equation}
with both top quarks decaying leptonically. As per Ref.~\cite{Fuks:2024yjj}, we
project out the colour singlet contribution to the process in
Eq.~\eqref{eq:toponium_mg5} and reweight the corresponding matrix elements
$\mathcal{M}$ by the ratio of the respective Green's function in the presence
and absence of a tree-level Coulomb potential,
\begin{equation}
    |\mathcal{M}|^2 \rightarrow |\mathcal{M}|^2 \left(\left|\frac{\tilde{G}(E;p^*)}{\tilde{G}_0(E;p^*)}\right|^2 - 1\right) \, , 
    \label{eq:toponium_reweighting}
\end{equation}
where $E$ and $p*$ denote respectively the toponium's binding energy, and the
momentum of the top and anti-top quark viewed from the toponium's rest frame.
The $-1$ term in Eq.~\eqref{eq:toponium_reweighting} serves to subtract the LO
contribution that is already present in the perturbative SM $t\bar{t}$
events.
As remarked in
Ref.~\cite{Fuks:2024yjj}, for $m_{t\bar{t}} > 350$ GeV the average top quark
velocity becomes too large to ensure non-relativistic kinematics, and we
therefore apply generator-level cuts at $m_{t\bar{t}}<350$ GeV and $p^*<50$ GeV. Other ways of simulating toponium based on
simplified models exist~\cite{Fuks:2021xje, Maltoni:2024tul}, such as through a
pseudo-scalar $^1S_0$. However, given that these neglect total angular momentum
contributions beyond $J=0$ we shall not consider these further in the current
work.

\begin{figure}[t]
    \centering
    \includegraphics[width=0.45\linewidth]{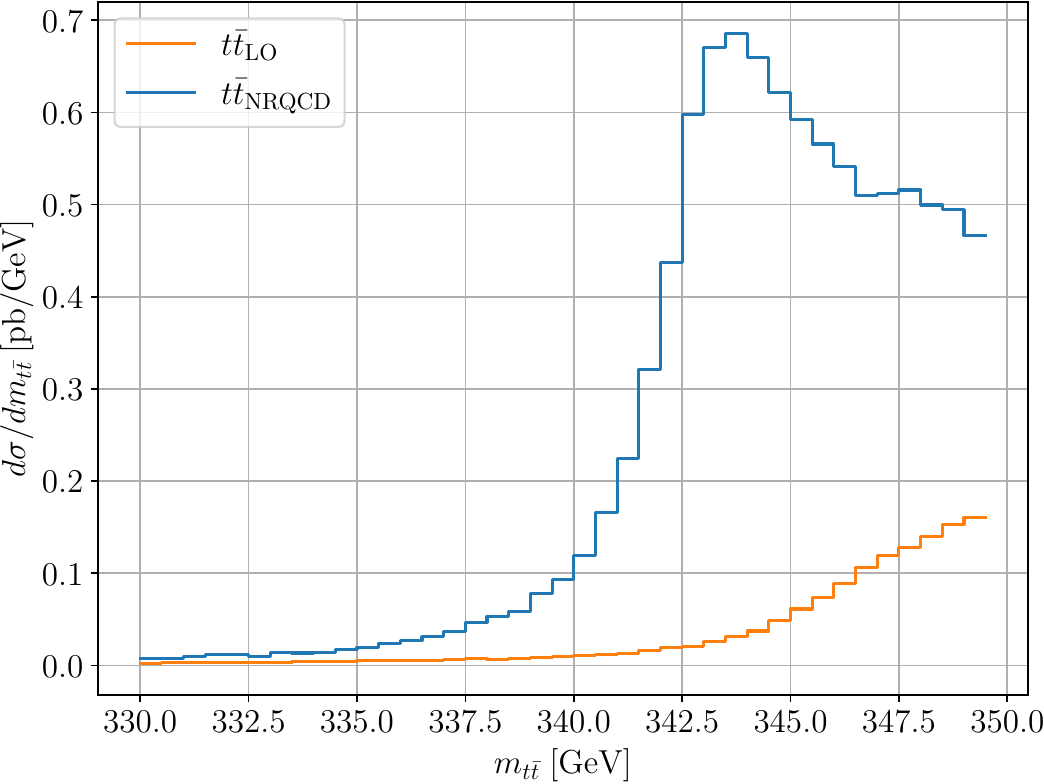}
    \includegraphics[width=0.45\linewidth]{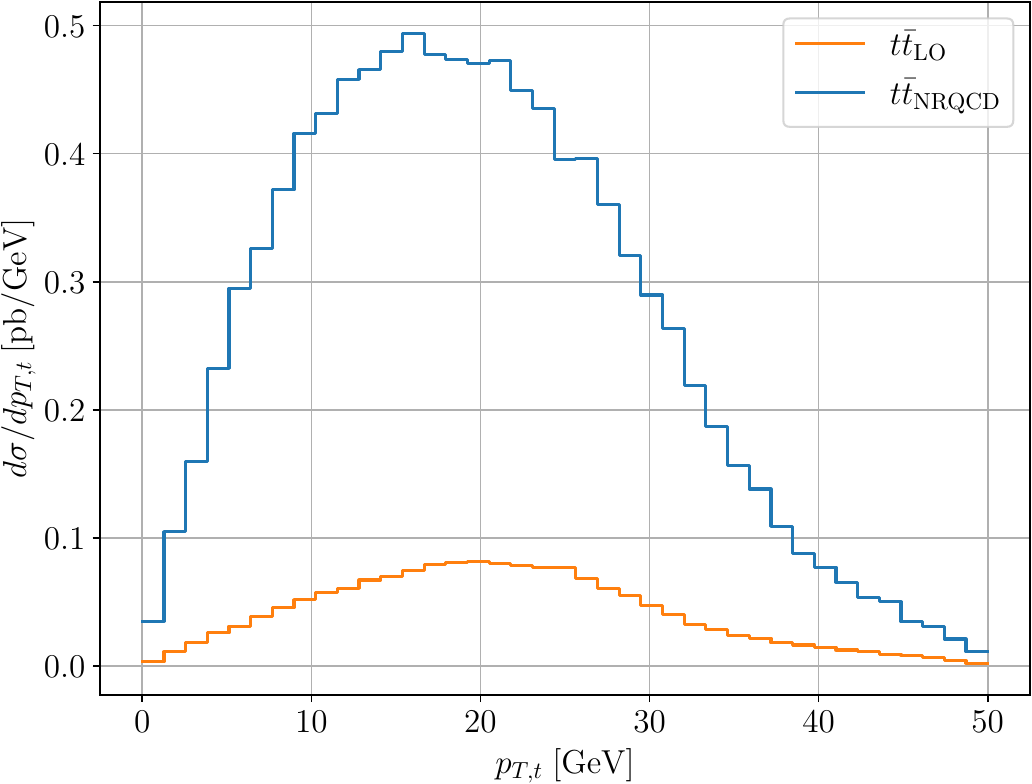}
    \caption{Differential distributions in the invariant mass $m_{t\bar{t}}$
    (left panel) and top-quark transverse momentum $p_T^t$ (right panel) at LO
    comparing the toponium signal ($t\bar{t}_{\rm NRQCD}$) modelled with NRQCD
    following the reweighting from Eq.~\eqref{eq:toponium_reweighting}
    against the nominal $t\bar{t}$ background at 13 TeV while imposing
    $m_{t\bar{t}} < 350$ GeV. 
    \label{fig:toponium-mtt-pT-dist}
    }
\end{figure}

Fig.~\ref{fig:toponium-mtt-pT-dist} displays the impact of the reweighting
according to Eq.~\eqref{eq:toponium_reweighting} on the differential
cross-section in the invariant mass $m_{t\bar{t}}$ (left panel) and the
transverse momentum $p_T^t$ (right panel), comparing the reweighted events,
labelled $t\bar{t}_{\rm NRQCD}$, against the nominal LO QCD background (labelled
$t\bar{t}_{\rm LO}$) when no reweighting is applied. The toponium signal appears most
prominently as an increased contribution to the production cross-section around
$m_{t\bar{t}} = 2m_t - 2$ GeV, while its effect on $p_T^t$ is limited to
relatively low $p_T$. Its impact on the rapidity distribution rescales all bins
uniformly and is therefore not shown. In generating
Fig.~\ref{fig:toponium-mtt-pT-dist}, we used the
\verb|NNPDF40_nnlo_as_01180| PDF set~\cite{NNPDF:2021njg} and set the pole mass
to $m_t
= 172.5$ GeV. We furthermore note that the sample has been rescaled to the total
cross sections reported in Table I of Ref.~\cite{Fuks:2021xje} in order to
remove the effect of the leptonic branching ratios, to account for the
colour-octet contribution to Eq.~\eqref{eq:toponium_mg5}, and to allow for a
smooth interpolation between the threshold and relativistic
region~\cite{Sumino:2010bv}. 

In order to incorporate the toponium signal into our fit, we include it as an
additional contribution to our NNLO cross-section predictions in the absence of
toponium as described in Sect.~\ref{subsec:theory_settings}. In particular, we
multiply our NNLO QCD $t\bar{t}$ prediction in bin $i$ without toponium, denoted
$T_i^{t\bar{t}}$, by a toponium $k$-factor, 
\begin{equation}
     k_i^{\rm (topo)} \equiv 1 + T_i^{t\bar{t}_{\rm NRQCD}}/T_i^{t\bar{t}} \, 
     \label{eq:topo_kfact}
\end{equation}
where $T_i^{t\bar{t}_{\rm NRQCD}}$ denotes the toponium contribution obtained
from applying the reweighting from Eq.~\eqref{eq:toponium_reweighting}. The
toponium $k$-factors for the measurements listed in Table \ref{tab:datasets} are
provided in App.~\ref{sec:app-sup-res}. 

In this context, it is relevant to point out a subtlety regarding the perturbative matching between the Green's function method and the fixed order NNLO QCD predictions. As discussed in Refs.~\cite{Braun:1968njz,Melnikov:2014lwa,Eides:2014swa,Beneke:2016jpx}, and more recently in Ref.~\cite{Nason:2025hix}, one needs to be careful not to double count terms that are present in both the perturbative QCD expansion and the resummation of the Coulomb interactions to all orders. Specifically, the latter resums terms in the $(\alpha_s/v)$ expansion, where $v$ denotes the velocity of the top-quark in the toponium's rest frame, so that the $(\alpha_s/v)$ and $(\alpha_s/v)^2$ contributions already included in the NNLO QCD calculation must be subtracted consistently when combined. A quantitative assessment of the size of this overlap was computed in Ref.~\cite{Nason:2025hix} in the case of the CMS measurement from Ref.~\cite{CMS:2018htd}, where they predict a cross-section of $0.602\,{\rm pb/GeV}$ at pure NNLO QCD, while the threshold corrected result comes out slightly higher at $0.661\, {\rm pb/GeV}$. Correcting for the bin width then gives a difference between both calculations of $3.54$ pb. 

Motivated by this theoretical effect, and by the higher values of the toponium cross-section measured by ATLAS and CMS ~\cite{ATLAS:2025mvr,CMS:2025kzt}, we conservatively assign a fully correlated 50\% uncertainty to the toponium $k$-factor in Eq.~\eqref{eq:topo_kfact}, such that we can accommodate all the various uncertainties within one sigma. This is done by adding a further contribution to the fit covariance matrix,
\begin{equation}
    C_{ij}^{({\rm topo})} = \frac{1}{4}\,T_i^{t\bar{t}_{\rm NRQCD}}T_j^{t\bar{t}_{\rm NRQCD}} \, ,
    \label{eq:topo_covmat}
\end{equation}
which has a single non-vanishing eigenvalue as expected for a normalization uncertainty. 

\section{Results}
\label{sec:results}
We now present the main results of this work. First, in
Sect.~\ref{subsec:closure-test-results}, we validate our methodology in a
controlled setup with synthetic data following the closure testing
described in Sect.~\ref{subsec:closure-test}. We then move to applications to
real data, always including MHOUs, presenting first results at NNLO QCD accuracy
in Sect.~\ref{subsec:as_mt_nnlo}. Here we also study in detail the impact of
different ways of statistically combining multiple measurements. We then study a
series of refinements, starting with a${\rm N^3LO}$ and a${\rm N^3LO}\otimes
\rm{NLO}_{\rm QED}$ corrections in Sect.~\ref{subsec:as_mt_n3lo}. The impact of
toponium corrections are studied in Sect.~\ref{subsec:topo_results}, while the
impact of constraining the prior on $\alpha_s(m_Z)$ to the recent FLAG
determination~\cite{FlavourLatticeAveragingGroupFLAG:2024oxs} is presented in
Sect.~\ref{subsec:flag_results}. We end in Sect.~\ref{subsec:comp_lit} by
comparing our results to other $m_t$ determinations in the literature.

The fits in this section are all performed in exactly the same way: we choose a 
global dataset (either real data or, in Sect.~\ref{subsec:closure-test} pseudo-data), 
including a selection of the available top data, add the theory covariance matrix $S$
computed according to the chosen prior in $(m_t,\alpha_s)$ as set out in Sect.~\ref{subsec:tcm_method} and Sect.~\ref{subsec:theory_settings} to the usual covariance matrix (which itself includes nuclear uncertainties and MHOU in addition to the experimental uncertainties), and then perform a global PDF fit using the usual NNPDF4.0 methodology~\cite{NNPDF:2021njg} through the NNPDF framework~\cite{NNPDF:2021uiq}. We then simply determine the posterior distribution of $(m_t,\alpha_s)$ using the formulae set out in Sect.~\ref{subsec:tcm_method}. If the shift from prior to posterior is particularly large, the fit can be repeated with a new prior, though in practice this is not usually necessary.

\subsection{Closure test results}
\label{subsec:closure-test-results}
\begin{figure}[ht!]
    \centering
    \includegraphics[width=0.6\linewidth]{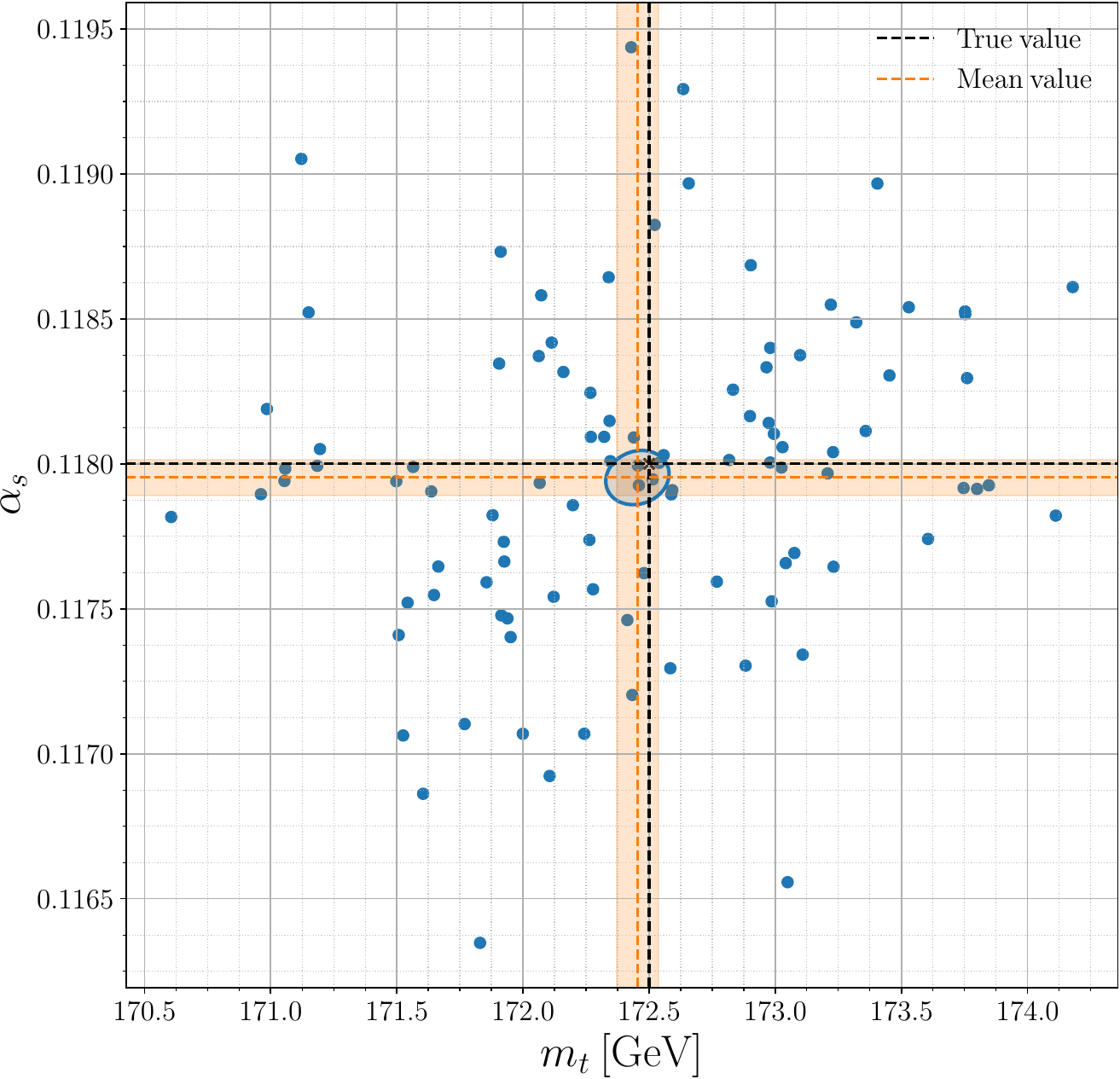}
    \caption{A closure test of $\alpha_s(m_Z)$ and $m_t$ generated from 100 $L_1$ instances, each consisting of 100 $L_2$ replicas, at NNLO QCD while including MHOUs with underlying truth $\alpha_s^*=0.118$ and $m_t^*=172.5$ GeV. The weighted average of the $L_1$ instances, indicated by the (blue) ellipse, is compatible with the true values at 68\% CL.}
    \label{fig:closure-test}
\end{figure}

Fig.~\ref{fig:closure-test} displays the result of a closure test, performed at
$\alpha_s^*=0.118$ and $m_t^*= 172.5$~GeV at NNLO QCD accuracy, including MHOUs.
In total, we analyse $N_{L_1}=100$ replicas, each in turn consisting of
$N_{L_2}=100$ replicas, and show how their central values are distributed around
the true values $\alpha_s^*$ and $m_t^*$. The corresponding weighted average and
its covariance, Eqs.~\eqref{eq:mean_closure} and \eqref{eq:cov_closure}, are
indicated by the (blue) ellipse, with the marginalised 68\% confidence level
(C.L.) intervals indicated by the (orange) shaded bands. We find compatibility
between truth and the reconstructed values within 1$\sigma$, and thus conclude
that our methodology is free from any significant bias. With only $N_{L_2}=100$ 
replicas, we are confident that our bootstrap uncertainties are significantly 
less than one per mille in both $\alpha_s^*$ and $m_t^*$.

Although the closure test should pass regardless of the specific data sets
included, we note for definiteness that the closure test shown in
Fig.~\ref{fig:closure-test} includes all $t\bar{t}$ data sets differential in
the top-quark pair rapidity $y_{t\bar{t}}$ as specified in Table
\ref{tab:datasets}. This is motivated by the fact that in our fits to real data,
to be described in the next section, it is precisely this observable that
is the least useful for determining $m_t$, producing results not always consistent 
with the more precise results given by some of the other observables. It is therefore interesting to see that even so, this observable still passes the closure test, despite these difficulties. 

Near kinematic boundaries, where cross-sections are small, our assumption that
data are Gaussianly distributed may lead to a small fraction of
negative replicas. In order to remedy this, the NNPDF4.0 determination imposed a
constraint through the introduction of a Lagrange
multiplier that penalised replicas giving negative cross-sections~\cite{NNPDF:2021njg}. 
However, based on
the fact that the TCM formalism assumes all data is Gaussian, we drop this constraint in the current work. However we continue to impose positivity on the PDFs themselves in the $\overline{\rm MS}$-scheme in the perturbative region~\cite{Candido:2023ujx}. This approach 
is slightly different to that adopted in the recent $\alpha_s$ determination
in Ref.~\cite{Ball:2025xgq} where PDF positivity was also dropped in the closure test.

\subsection{Determining $m_t$ and $\alpha_s(m_Z)$ at NNLO}
\label{subsec:as_mt_nnlo}

Having validated our methodology in Sect.~\ref{subsec:closure-test-results} on
pseudo data, we now move to applications to real data. Recalling that
the various observables determined from the same data set are not independent, and that their
inter-spectra correlations are not always (publicly) available, we study in the
following the impact of each of the kinematic observables separately.
In the few cases in which inter-spectra correlations are available, we comment
on this explicitly and study their combined impact.

\begin{figure}[htbp]
    \centering
    \includegraphics[width=0.65\linewidth]{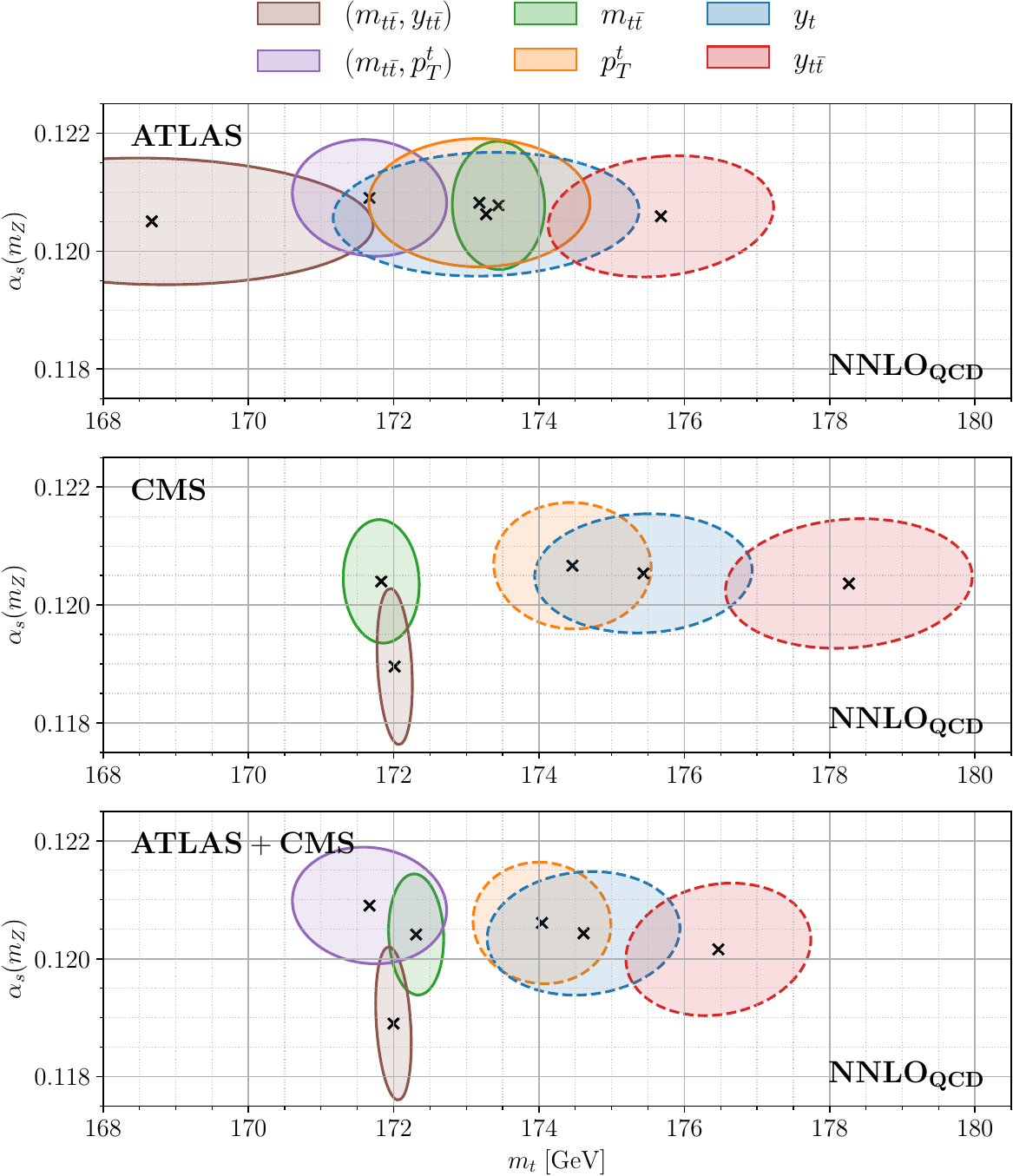}
    \caption{68\% C.L. bounds on the top-quark's mass $m_t$ and the strong
    couplings constant $\alpha_s(m_Z)$, at NNLO QCD with MHOUs. In each case,
    multiple exclusion contours are shown, each corresponding to a fit performed
    on the baseline dataset complemented with top-quark data differential in the
    given observable. Each fit is done with 500 replicas. Ellipses with dashed lines indicate that at least one of the data sets in this measurement has a 
  $\chi^2/n_{\rm dat} > 3.0$}
    \label{fig:obs-comparison-nnlo}
\end{figure}

\begin{table}[htbp]
  \centering
  \scriptsize
  \renewcommand{\arraystretch}{1.1}
  \begin{tabular}{lllll}
\toprule
Observable & ATLAS & CMS & ATLAS + CMS \\
\midrule
\multirow{3}{*}{$(m_{t\bar{t}}, p_T^t)$} & $m_t = 171.67 \pm 0.70$ &  &  \\
 & $\alpha_s = 0.12090 \pm 0.00066$ &  &  \\
 & $\rho = -0.082$ &  &  \\
 \midrule
\multirow{3}{*}{$(m_{t\bar{t}}, y_{t\bar{t}})$} & $m_t = 168.67 \pm 2.02$ & $m_t = 172.01 \pm 0.16$ & $m_t = 172.00 \pm 0.16$ \\
 & $\alpha_s = 0.12051 \pm 0.00071$ & $\alpha_s = 0.11896 \pm 0.00088$ & $\alpha_s = 0.11890 \pm 0.00086$ \\
 & $\rho = -0.063$ & $\rho = -0.241$ & $\rho = -0.262$ \\
 \midrule
\multirow{3}{*}{$m_{t\bar{t}}$} & $m_t = 173.44 \pm 0.42$ & $m_t = 171.83 \pm 0.35$ & $m_t = 172.31 \pm 0.25$ \\
 & $\alpha_s = 0.12078 \pm 0.00072$ & $\alpha_s = 0.12040 \pm 0.00069$ & $\alpha_s = 0.12041 \pm 0.00068$ \\
 & $\rho = -0.016$ & $\rho = -0.054$ & $\rho = -0.072$ \\
 \midrule
\multirow{3}{*}{$p_T^t$} & $m_t = 173.18 \pm 1.01$ & $m_t = 174.46 \pm 0.72$ & $m_t = 174.04 \pm 0.63$ \\
 & $\alpha_s = 0.12082 \pm 0.00072$ & $\alpha_s = 0.12067 \pm 0.00071$ & $\alpha_s = 0.12061 \pm 0.00068$ \\
 & $\rho = -0.003$ & $\rho = -0.026$ & $\rho = -0.040$ \\
 \midrule
\multirow{3}{*}{$y_t$} & $m_t = 173.27 \pm 1.39$ & $m_t = 175.44 \pm 0.99$ & $m_t = 174.61 \pm 0.88$ \\
 & $\alpha_s = 0.12063 \pm 0.00070$ & $\alpha_s = 0.12054 \pm 0.00067$ & $\alpha_s = 0.12043 \pm 0.00069$ \\
 & $\rho = 0.060$ & $\rho = 0.054$ & $\rho = 0.088$ \\
 \midrule
\multirow{3}{*}{$y_{t\bar{t}}$} & $m_t = 175.68 \pm 1.03$ & $m_t = 178.27 \pm 1.12$ & $m_t = 176.47 \pm 0.84$ \\
 & $\alpha_s = 0.12059 \pm 0.00068$ & $\alpha_s = 0.12036 \pm 0.00073$ & $\alpha_s = 0.12016 \pm 0.00074$ \\
 & $\rho = 0.151$ & $\rho = 0.108$ & $\rho = 0.146$ \\
\bottomrule
\end{tabular}

  \vspace{0.3cm}
  \caption{The 68 \% C.L. intervals on the top-quark mass, $m_t$, strong
  coupling constant, $\alpha_s$, and their correlation coefficient, $\rho$, at NNLO with
  MHOUs, for each of the observables plotted in Fig.~\ref{fig:obs-comparison-nnlo}, shown
  separately for ATLAS, CMS, and their combination (ATLAS+CMS). }
  \label{tab:mt_alphas_NNLO_no_topo}
\end{table}

Fig.~\ref{fig:obs-comparison-nnlo} displays the $68\%$ C.L. intervals at NNLO
QCD accuracy in the ($\alpha_s, m_t$) plane corresponding to fits carried out on
our baseline data set, each supplemented with $t\bar{t}$ measurements
differential in a variety of different kinematic observables. The numerical central values, uncertainties 
and correlation coefficients for the various ellipses are provided in Table \ref{tab:mt_alphas_NNLO_no_topo}.
In total, we consider measurements differential either in the invariant mass
$m_{t\bar{t}}$ of the top-quark pair, the top-quark's transverse momentum $p_T^t$, 
the top-quark's rapidity $y_t$, the rapidity of the $t\bar{t}$ pair, $y_{t\bar{t}}$, 
and then also double differential distributions in either $(m_{t\bar{t}}, p_T^t)$ or
$(m_{t\bar{t}}, y_{t\bar{t}})$. From top to bottom, we show first a
determination based only on ATLAS $t\bar{t}$ measurements, followed by CMS $t\bar{t}$
measurements in the middle panel, and finally their combination in the bottom
panel. All determinations share exactly the same data sets except for the relevant
$t\bar{t}$ measurements. We refer to Table \ref{tab:chi2-no-toponium} for an
overview of the $t\bar{t}$ data sets that enter each determination, together
with their $\chi^2$ normalised to the number of data points $n_{\rm dat}$. In
Fig.~\ref{fig:obs-comparison-nnlo}, ellipses with dashed lines indicate that at
least one of the associated data sets has a $\chi^2/n_{\rm dat} > 3.0$.

\begin{sidewaystable}
  \centering
  \scriptsize
  \renewcommand{\arraystretch}{1.1}
 
  \begin{tabular}{lll|ccc|ccc|ccc}
\multicolumn{3}{c}{} &
\multicolumn{3}{c}{\textbf{ATLAS}} &
\multicolumn{3}{c}{\textbf{CMS}} &
\multicolumn{3}{c}{\textbf{ATLAS+CMS}} \\
Observable & Data set & \makecell{$n_{\rm dat}$} &
\makecell{NNLO\\QCD} &
\makecell{aN$^3$LO \\ QCD} & 
\makecell{aN$^3$LO$_{\rm QCD}$\\$\otimes$NLO$_{\rm QED}$} &
\makecell{NNLO\\QCD} &
\makecell{aN$^3$LO \\ QCD} & 
\makecell{aN$^3$LO$_{\rm QCD}$\\$\otimes$NLO$_{\rm QED}$} &
\makecell{NNLO\\QCD} &
\makecell{aN$^3$LO \\ QCD} & 
\makecell{aN$^3$LO$_{\rm QCD}$\\$\otimes$NLO$_{\rm QED}$} \\
\toprule
\multirow[t]{7}{*}{$m_{t\bar{t}}$} & ATLAS 13~TeV $t\bar{t}$ all hadr. & 9 & 1.180 & 1.266 & 1.264 & — & — & — & 1.088 & 1.142 & 1.135 \\
 & ATLAS 13~TeV $t\bar{t}$ $\ell+j$ & 9 & 0.955 & 0.944 & 0.883 & — & — & — & 0.932 & 0.907 & 0.850 \\
 & ATLAS 8~TeV $t\bar{t}$ $2\ell$ & 6 & 0.146 & 0.169 & 0.145 & — & — & — & 0.101 & 0.113 & 0.096 \\
 & ATLAS 8~TeV $t\bar{t}$ $\ell+j$  & 7 & 0.224 & 0.187 & 0.211 & — & — & — & 0.304 & 0.261 & 0.287 \\
 & CMS 13~TeV $t\bar{t}$ $2\ell$ 138 ${\rm fb}^{-1}$ & 7 & — & — & — & 1.566 & 1.690 & 1.629 & 1.569 & 1.702 & 1.640 \\
 & CMS 13~TeV $t\bar{t}$ $\ell+j$ & 15 & — & — & — & 0.902 & 0.975 & 0.990 & 0.912 & 0.992 & 1.000 \\
 & total & 31 & 0.856 & 0.852 & 0.843 & 1.186 & 1.237 & 1.226 & 1.160 & 1.170 & 1.163 \\
\cline{1-12}
\multirow[t]{2}{*}{$(m_{t\bar{t}}, p_T^t)$} & ATLAS 13~TeV $t\bar{t}$ $\ell+j$ & 15 & 0.517 & 0.474 & 0.440 & — & — & — & — & — & — \\
 & total & 15 & 0.517 & 0.474 & 0.440 & — & — & — & — & — & — \\
\cline{1-12}
\multirow[t]{4}{*}{$(m_{t\bar{t}}, y_{t\bar{t}})$} & ATLAS 13~TeV $t\bar{t}$ all hadr. & 11 & 2.561 & 2.707 & 2.850 & — & — & — & 1.182 & 1.102 & 1.121 \\
 & CMS 13~TeV $t\bar{t}$ $2\ell$ 138 ${\rm fb}^{-1}$ & 16 & — & — & — & 1.789 & 1.732 & 1.766 & 1.765 & 1.682 & 1.684 \\
 & CMS 13~TeV $t\bar{t}$ $\ell+j$ & 34 & — & — & — & 2.461 & 2.428 & 2.467 & 2.435 & 2.347 & 2.374 \\
 & total & 11 & 2.561 & 2.707 & 2.850 & 2.342 & 2.304 & 2.343 & 2.132 & 2.051 & 2.074 \\
\cline{1-12}
\multirow[t]{5}{*}{$p_T^t$} & ATLAS 13~TeV $t\bar{t}$ $\ell+j$ & 8 & 0.839 & 0.837 & 0.786 & — & — & — & 0.814 & 0.815 & 0.769 \\
 & ATLAS 8~TeV $t\bar{t}$ $\ell+j$  & 8 & 0.371 & 0.337 & 0.309 & — & — & — & 0.382 & 0.345 & 0.315 \\
 & CMS 13~TeV $t\bar{t}$ $2\ell$ 138 ${\rm fb}^{-1}$ & 7 & — & — & — & 3.110 & 3.253 & 3.194 & 3.080 & 3.222 & 3.194 \\
 & CMS 13~TeV $t\bar{t}$ $\ell+j$ & 16 & — & — & — & 0.517 & 0.551 & 0.536 & 0.507 & 0.540 & 0.536 \\
 & total & 16 & 0.561 & 0.548 & 0.516 & 1.421 & 1.460 & 1.437 & 1.061 & 1.081 & 1.069 \\
\cline{1-12}
\multirow[t]{5}{*}{$y_t$} & ATLAS 13~TeV $t\bar{t}$ $\ell+j$ & 5 & 0.668 & 0.663 & 0.693 & — & — & — & 0.680 & 0.665 & 0.708 \\
 & ATLAS 8~TeV $t\bar{t}$ $\ell+j$  & 5 & 4.191 & 4.291 & 4.625 & — & — & — & 3.731 & 3.831 & 4.029 \\
 & CMS 13~TeV $t\bar{t}$ $2\ell$ 138 ${\rm fb}^{-1}$ & 10 & — & — & — & 3.162 & 3.129 & 3.109 & 3.168 & 3.119 & 3.099 \\
 & CMS 13~TeV $t\bar{t}$ $\ell+j$ & 11 & — & — & — & 3.164 & 3.261 & 3.325 & 2.980 & 3.064 & 3.052 \\
 & total & 10 & 2.549 & 2.600 & 2.790 & 3.088 & 3.129 & 3.151 & 2.809 & 2.822 & 2.844 \\
\cline{1-12}
\multirow[t]{7}{*}{$y_{t\bar{t}}$} & ATLAS 13~TeV $t\bar{t}$ all hadr. & 12 & 0.813 & 0.818 & 0.824 & — & — & — & 0.701 & 0.700 & 0.707 \\
 & ATLAS 13~TeV $t\bar{t}$ $\ell+j$ & 7 & 0.658 & 0.677 & 0.696 & — & — & — & 0.378 & 0.379 & 0.388 \\
 & ATLAS 8~TeV $t\bar{t}$ $2\ell$ & 5 & 1.075 & 1.064 & 1.126 & — & — & — & 0.720 & 0.686 & 0.720 \\
 & ATLAS 8~TeV $t\bar{t}$ $\ell+j$  & 5 & 3.741 & 3.656 & 3.908 & — & — & — & 2.341 & 2.220 & 2.446 \\
 & CMS 13~TeV $t\bar{t}$ $2\ell$ & 10 & — & — & — & 0.812 & 0.840 & 0.928 & 0.615 & 0.634 & 0.695 \\
 & CMS 13~TeV $t\bar{t}$ $\ell+j$ & 10 & — & — & — & 4.487 & 4.392 & 4.516 & 3.676 & 3.622 & 3.706 \\
 & total & 29 & 1.275 & 1.261 & 1.318 & 2.866 & 2.838 & 2.965 & 1.537 & 1.517 & 1.578 \\
\bottomrule
\end{tabular}

    \caption{The $\chi^2/n_{\rm dat}$ of each of the data sets entering the
  joint $\alpha_s(m_Z)$, $m_t$ determinations split up according to whether only
  data sets from ATLAS, CMS or both are used as input to the fit. In each case,
  the $\chi^2$ is given at NNLO QCD, aN$^3$LO QCD, and aN$^3$LO QCD with QED
  corrections. The corresponding
  bounds are provided in Table \ref{tab:mt_bounds_overview} and shown in
  Figs.~\ref{fig:obs-comparison-nnlo} and \ref{fig:obs-comparison-n3lo-qed}. In
  all cases MHOUs are included. \label{tab:chi2-no-toponium}}
 
\end{sidewaystable}

Several interesting observations can be made by inspecting
Fig.~\ref{fig:obs-comparison-nnlo} together with the $\chi^2$ values in Table
\ref{tab:chi2-no-toponium}. First of all, we note a significant spread in $m_t$
for the different ellipses, while $\alpha_s$ is instead relatively stable. This
can be understood by noting that the data sets common to all the fits,
in particular the DIS, Drell-Yan, single-jet and di-jet measurements, are 
themselves all sensitive to
$\alpha_s$, while having no sensitivity to $m_t$. Thus $\alpha_s$ is mainly 
determined by these data sets, leaving the different top-quark data sets to 
determine $m_t$. An exception is the determination based on the
$(m_{t\bar{t}}, y_{t\bar{t}})$ distribution from CMS, where we observe a rather
lower value of $\alpha_s(m_Z)$. We traced this effect to a slight tension
between this particular dataset and the single-jet and di-jet measurements.
It is important to remark
that the spread in central values of $m_t$ from the different observables has 
nothing to do with the fitting methodology, since it disappears once considered in a
closure test. Indeed, we refer to Fig.~\ref{fig:closure-test} for the equivalent
result of the $y_{t\bar{t}}$ contour from Fig.~\ref{fig:obs-comparison-nnlo} in
the context of a closure test. It is interesting to note that the outlying observables 
tend to be those with relatively poor $\chi^2$. That the spread is somewhat broader 
in the case of CMS suggests the spread in $m_t$ may in part be an experimental
issue. At the same time, to the extent ATLAS and CMS display a similar pattern suggests that there may also be 
theoretical issues, perhaps in the theoretical description of other datasets in the 
fit such as jets~\cite{Ball:2025xtj}. 

Second, we observe a strong hierarchy in the precision with which each observable
can be used to extract $m_t$, which is generally consistent between both ATLAS and CMS. The
$(m_{t\bar{t}}, y_{t\bar{t}})$ double differential distributions from CMS in the dilepton and $\ell +j$ channels provide the most stringent bounds on $m_t$, and these dominate in the combination.  The
$(m_{t\bar{t}}, y_{t\bar{t}})$ distribution from ATLAS consists only
of measurements in the fully hadronic channel which is plagued by large
experimental and theoretical uncertainties, thus providing only a poor extraction of $m_t$. 
It is interesting that when this distribution is fitted together with the CMS
measurements its $\chi^2$ improves significantly, from 2.56 to 1.18. The single
differential distributions in $m_{t\bar{t}}$, from both ATLAS and CMS, are all
well fitted, and give rather precise results for $m_t$, as expected from the
discussion in Fig.~\ref{fig:kinematic_dist}. The fact that the double
differential distribution outperforms the single differential measurement matches the naive expectation that integrating over one of the observables incurs a loss of information, which in
turn translates into a less precise $m_t$ determination. The $p_T^t$ distributions give somewhat less precise results for $m_t$, and one them is rather poorly fitted.

The observables least sensitive to $m_t$ are clearly the distributions in rapidities, 
either $y_t$ or $y_{t\bar{t}}$. 
This we anticipated already, based on the kinematic distributions shown in
Fig.~\ref{fig:kinematic_dist}, where varying $m_t$ mainly changes
the normalisation of the rapidity distribution rather than inducing any
noticeable shape effects. Furthermore, besides its large uncertainty in $m_t$,
the CMS $y_{t\bar{t}}$ distribution is a clear outlier compared to the other
extractions. This can be explained by 
inspecting its $\chi^2$ from Table
\ref{tab:chi2-no-toponium} -- the CMS 13 TeV measurement in the $\ell + j$
channel has a particularly high $\chi^2$ of 4.49 (though again this improves in 
the combined fit). We have verified that, for this particular data set, the fitted predictions lie consistently above the experimental data, which the TCM corrects for by an increased value of $m_t$. Difficulties in fitting individual rapidity distributions have been previously
encountered in Refs.~\cite{Bailey:2019yze, Cridge:2023ztj, NNPDF:2021njg, Ablat:2023tiy}.

Focusing now on the ATLAS+CMS combination, we note how precision improves with
respect to considering either ATLAS or CMS separately, as expected for mutually 
consistent data. Moreover, results
become increasingly more compatible when one restricts attention to determinations 
with a $\chi^2/n_{\rm dat} < 3.0$, with the $p_T^t$, $(m_{t\bar{t}}, y_{t\bar{t}})$ and $m_{t\bar{t}}$ all 
now overlapping within 1$\sigma$. The observable most sensitive to $m_t$ with a
satisfactory $\chi^2$ of 1.16 is the
single differential distribution in $m_{t\bar{t}}$, which
in the ATLAS+CMS combination results in $m_t = 172.31 \pm 0.25$ GeV.

\begin{figure}[t!]
    \centering
    \includegraphics[width=\linewidth]{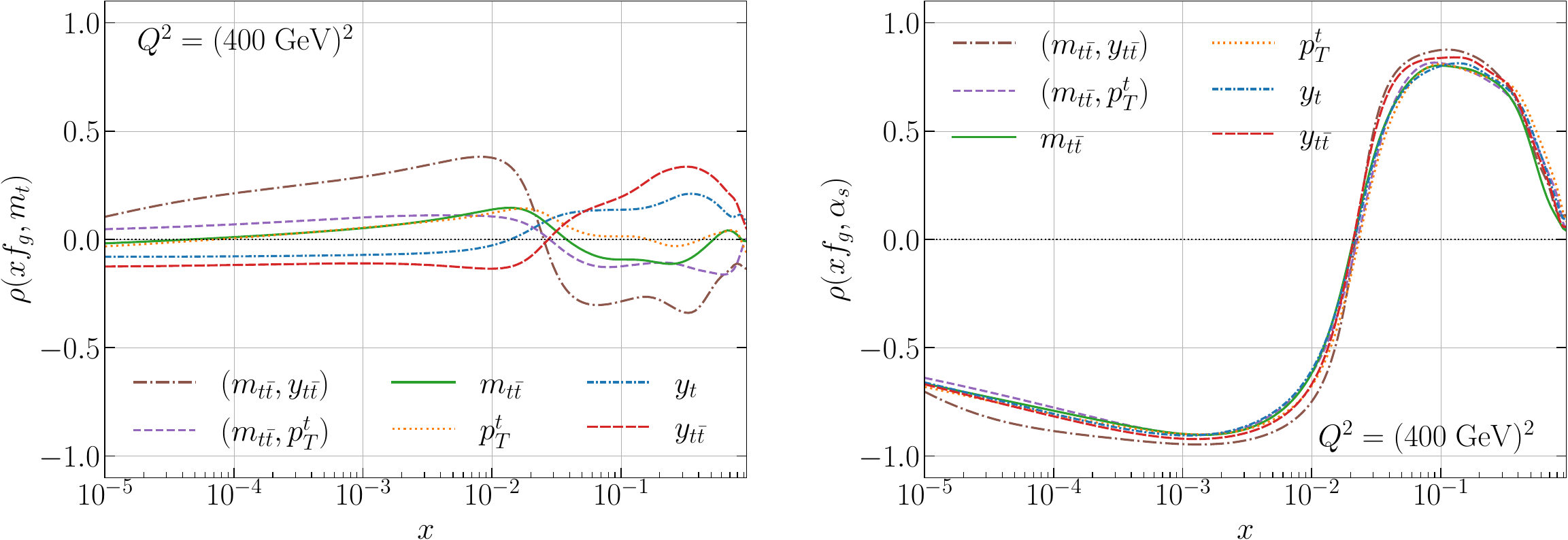}
    \caption{Left: The correlation $\rho(xf_g, m_t)$ between the gluon PDF $f_g$
    at NNLO QCD with MHOUs 
    and the top-quark mass $m_t$ as a function of Bjorken-$x$, for each of the
    observables in Fig.~\ref{fig:obs-comparison-nnlo}. Right: same, but
    for $\alpha_s$ instead of $m_t$. }
    \label{fig:correlation-pdf-mt-alphas}
\end{figure}

Regarding correlations between $m_t$ and $\alpha_s(m_Z)$, we find that for
$m_{t\bar{t}}$ and $p_T^t$ they are consistently negative, while for rapidities
they are consistently positive. This can be intuitively understood by noting that the rapidity is most
sensitive to the gluon PDF which in turn affects $t\bar{t}$ production through
the gluon fusion channel. This is consistent with Fig.~\ref{fig:correlation-pdf-mt-alphas}, which displays the correlation pattern between the gluon
PDF $f_g$ and $m_t$ (left panel), and between $f_g$ and $\alpha_s(m_Z)$ (right panel), as a function of
Bjorken-$x$ for each of the observables considered in
Fig.~\ref{fig:obs-comparison-nnlo}. We find that wherever the rapidities show positive
correlations between $m_t$ and the gluon PDF, they also show positive correlations
with $\alpha_s(m_Z)$,
while the opposite effect is instead observed for the other observables. The size of the correlations also reflects the degree
of correlation between $\alpha_s(m_Z)$ and $m_t$. In the $x$-range relevant for
$t\bar{t}$ production, the correlation of the gluon PDF with $m_t$ is positive for
$y_{t\bar{t}}$, less positive for $y_t$, but negative for $m_{t\bar t}$ and the joint distribution
$(m_{t\bar{t}}, y_{t\bar{t}})$, consistent with the correlation pattern
in Table \ref{tab:mt_alphas_NNLO_no_topo}. 

Further inspecting the correlations in Fig.~\ref{fig:correlation-pdf-mt-alphas},
we find that the gluon PDF is much more correlated to $\alpha_s(m_Z)$ than
$m_t$, which is consistent with the fact that $\alpha_s$ enters in the evolution and hard cross-sections of all the data included in the fit,
while $m_t$ only enters the hard cross-sections of the top datasets. It follows that the correlation of the gluon with $\alpha_s$ is insensitive to the choice of top dataset, since it is predominantly due to the rest of the data in the global fit. While for $\alpha_s(m_Z)$ we
find correlations of up to 0.75 in the mid to high-$x$ region, for $m_t$ the correlation is still as much 0.40, which is still significant. The change in sign of the correlation of $\alpha_s$ with the gluon (a consequence of the momentum sum rule) is reflected in a similar change of sign in the correlation of $m_t$ with the gluon.

These results show that while the inclusion of correlations with the PDFs
through the TCM (in what amounts to a joint fit) is essential for a global
determination of $\alpha_s$~\cite{Ball:2025xgq, Ball:2025xtj}, it is also
necessary for a reliable determination of the top mass $m_t$. The correlation of
$m_t$ with $\alpha_s$ is also significant, but rather less, and can be minimised
by using the differential $m_{t\bar{t}}$ and $p_T^t$ distributions. This gives
us confidence that our best results for 
$m_t$ are quite insensitive to the value of $\alpha_s$, which is in agreement with the observations of Ref.~\cite{Cridge:2023ztj}, being determined rather by the shape of the top observables.

\paragraph{Statistical combinations}

\begin{figure}[htbp]
    \centering
    \includegraphics[width=\linewidth]{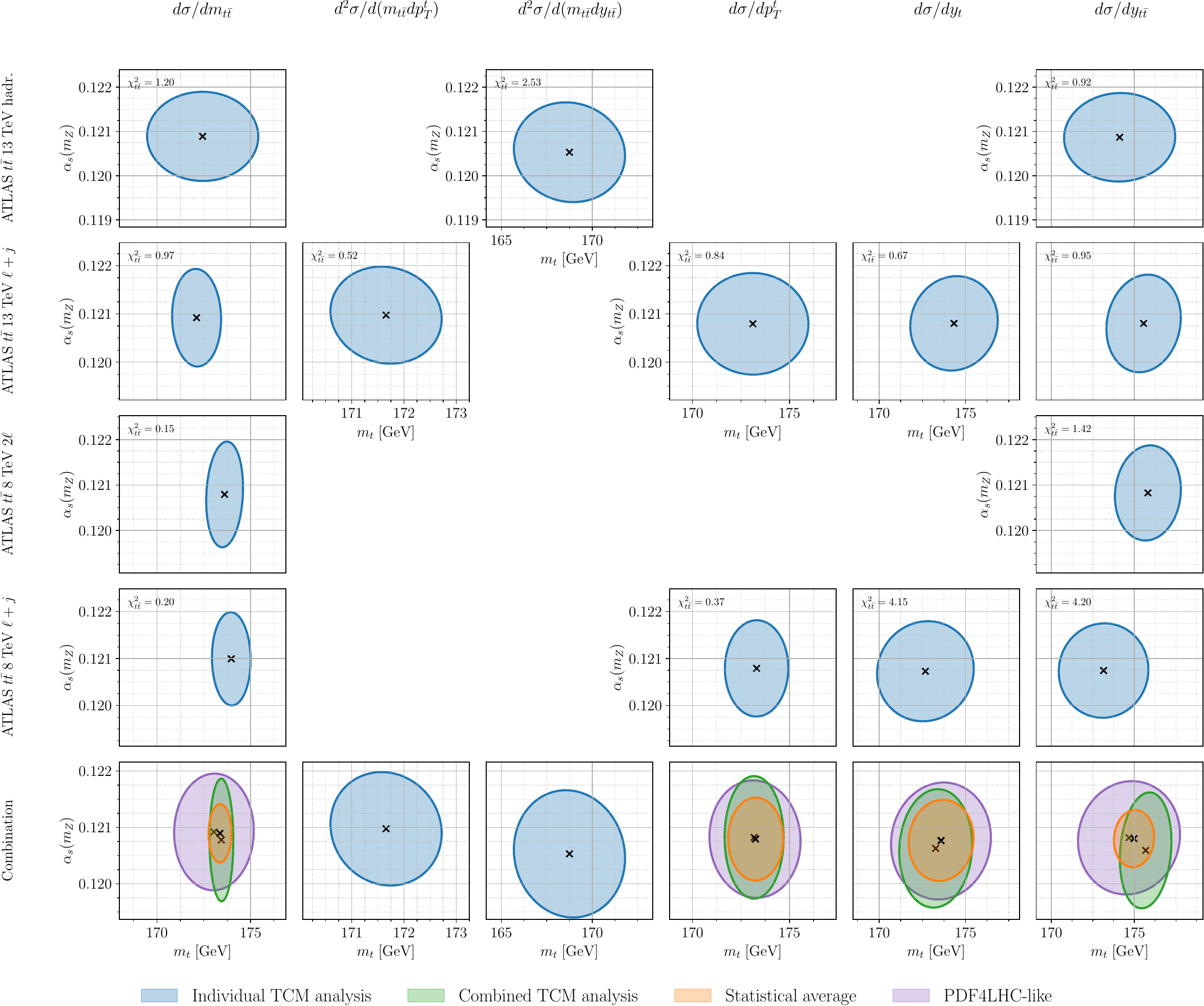}
    \caption{68 \% C.L. contours on the top-quark's mass $m_t$ and the strong coupling constant $\alpha_s(m_Z)$ at NNLO QCD with MHOUs obtained from separate fits to top-quark data differential in distinct kinematic observables (columns) for various ATLAS data sets (rows), whenever available (all in blue). The bottom row shows the combination of all the datasets in the corresponding column, for each observable, made three different ways: the combined TCM analysis (in green), corresponding to a simultaneous fit including all data sets within each column, the statistical average of the individual TCM ellipses (in orange), and finally, the PDF4LHC-like combination obtained by combining all replicas with equal weights (in purple).}
    \label{fig:dataset-breakdown-atlas}
\end{figure}

\begin{figure}[htbp]
    \centering
    \includegraphics[width=\linewidth]{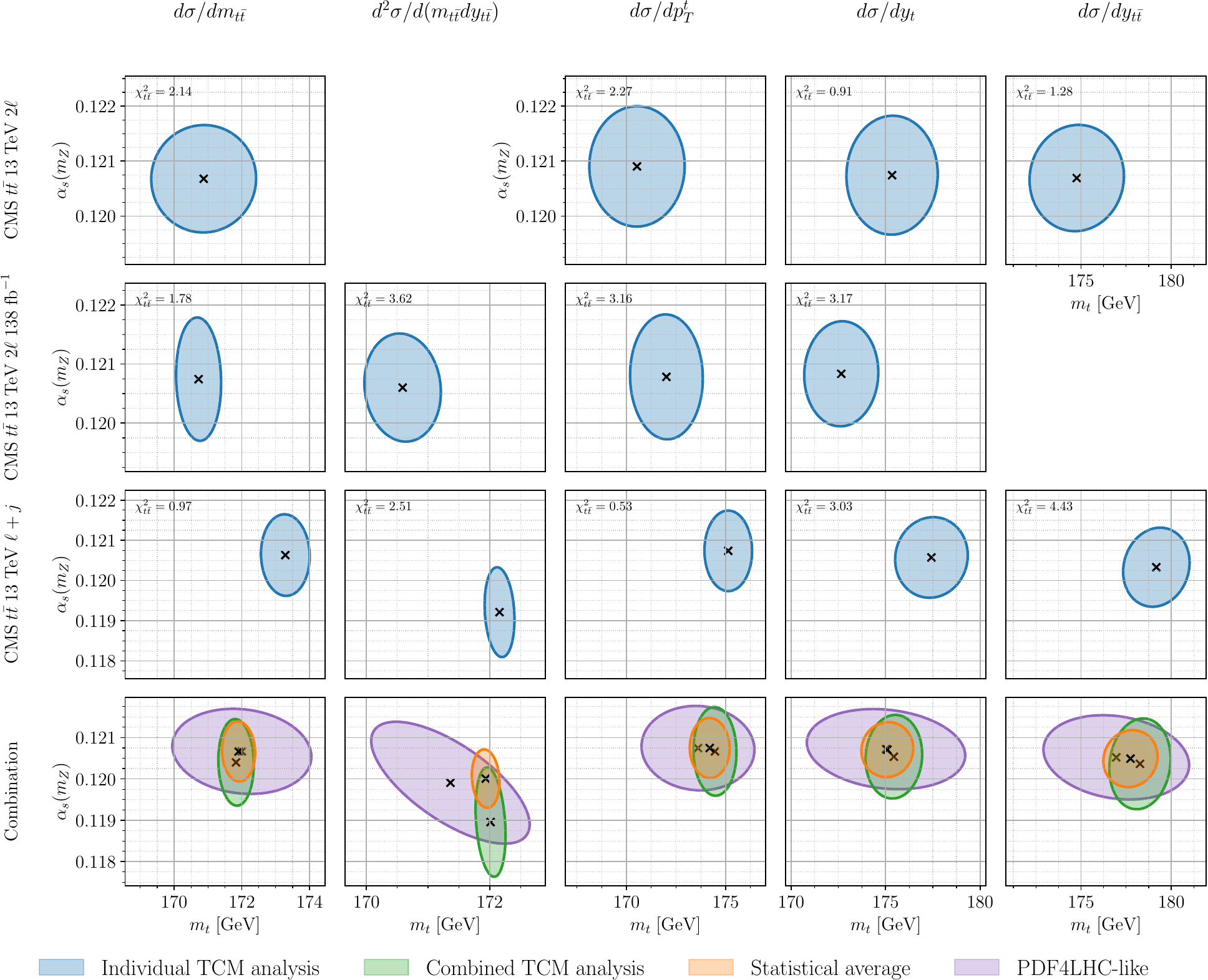}
    \caption{Same as Fig.~\ref{fig:dataset-breakdown-atlas}, now for the CMS
    datasets.}
    \label{fig:dataset-breakdown-cms}
\end{figure}

An important question concerns how to statistically combine multiple
measurements in order to faithfully account for all sources of uncertainties. This is a topic much discussed, both within the PDG, but also in the context of PDFs (see, for example,
Ref.~\cite{Ball:2021dab} and references therein). To
this end, we show in Figs.~\ref{fig:dataset-breakdown-atlas} and
\ref{fig:dataset-breakdown-cms} the breakdown of the individual measurements
that enter Fig.~\ref{fig:obs-comparison-nnlo} in the case of ATLAS and CMS,
respectively. In each column, we display results for a single kinematic
observable, while the rows distinguish between the various data sets. Whenever a
given data set does not provide measurements differential in the given
observable, a blank entry is shown. In the bottom row, we display the
combination of all the data sets in the corresponding column, for each
observable, made in three different ways. First we show the combined TCM
analysis, which
corresponds to what is shown in Fig.~\ref{fig:obs-comparison-nnlo}. Its result
is obtained from a single fit including all data sets in this column.
Second, we show the statistical average, which does not correspond to
a new fit, but rather to an a posteriori combination of the previous ellipses
weighted according to their means and covariances following
Eqs.~\ref{eq:mean_closure} and \ref{eq:cov_closure}. The third way of combining
determinations is labelled ``PDF4LHC-like'', which corresponds to
instead an unweighted average where all replicas that make up the individual TCM
analyses are used collectively to determine a new exclusion contour~\cite{Butterworth:2015oua}. 

A clear pattern arises from the combinations in
Fig.~\ref{fig:dataset-breakdown-atlas} and Fig.~\ref{fig:dataset-breakdown-cms}. As expected, the ``PDF4LHC-like'' combination is the
most conservative, significantly overestimating uncertainties, (though always including the correct central value within this uncertainty). This is because it treats all the datasets with equal weight, irrespective of their precision or consistency. An envelope prescription would give even broader uncertainties.
The statistical average, by contrast, assumes that the determinations made with each top 
dataset are uncorrelated, ignoring the fact that apart from the top data, the underlying dataset used to determine the PDFs is the same in each. It thus underestimates uncertainties, especially in $\alpha_s$. In effect such a combination
essentially reuses the same data multiple times. Since 
we have considerably more data to constrain $\alpha_s$, such as all DIS, single-jet and
di-jet data, than we have for $m_t$, the effect is much more visible there. 
However it is also visible in the $m_t$ determination, particularly for the rapidity 
distributions where the correlation with the PDFs is stronger.
The correct combination is provided by the combined TCM analysis, which takes
into account properly all correlations between the data sets due to the
PDFs, $\alpha_s$ and $m_t$, with no double counting. 

\begin{figure}[htbp]
    \centering
    \includegraphics[width=.6\linewidth]{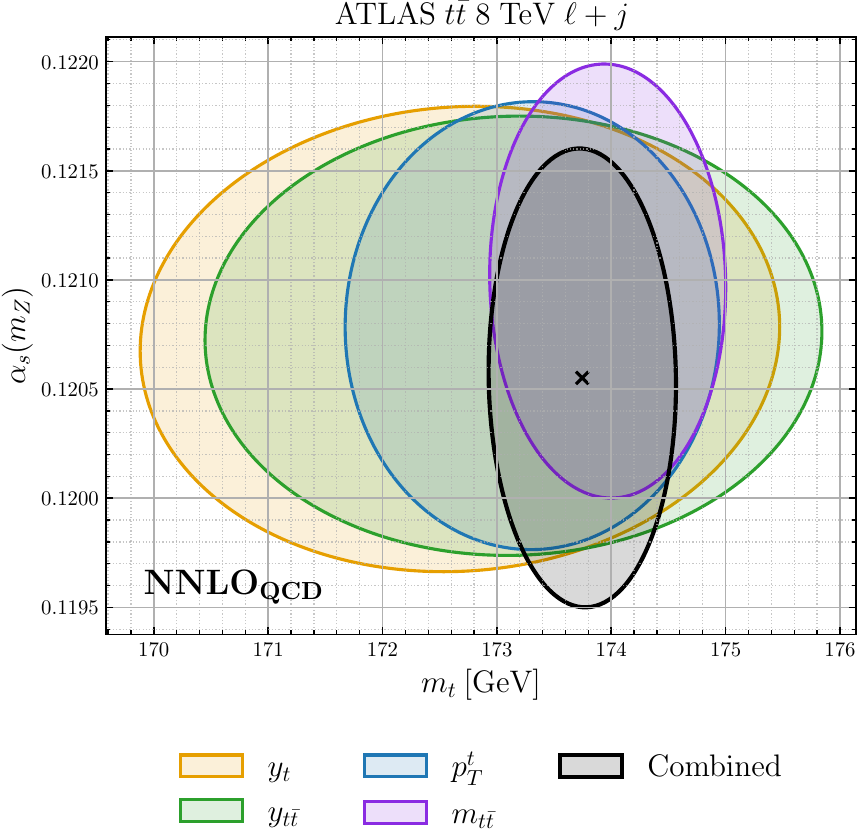}
    \caption{The 68\% C.L. contours in the ($\alpha_s, m_t$) plane comparing the
   impact of analysing each of the observables $y_t$, $y_{t\bar{t}}$, $p_T^t$,
   $m_{t\bar{t}}$ entering the ATLAS $t\bar{t}$ 8 TeV $\ell + j$ data set
   individually versus their combined impact when considering their
   inter-spectra correlations.}
    \label{fig:interspectra}
\end{figure}

\paragraph{Inter-spectra correlations} In the case of the ATLAS $t\bar{t}$ 8 TeV
measurement in the $\ell +j$ channel, inter-spectra correlations are known. This
means that all observables within this data set may be analysed simultaneously.
We display in Fig.~\ref{fig:interspectra} a comparison between the constraints
at the 68\% C.L. obtained either when analysing one observable at the time
versus the combination that considers $y_t$, $y_{t\bar{t}}$, $p_T^t$ and
$m_{t\bar{t}}$ simultaneously, and provide the corresponding $\chi^2$ values in
Table \ref{tab:chi2_interspectra_atlas_8tev_lj}. It is clear how the combination provides more
stringent constraints on $m_t$ than any of the observables
individually. Inspecting the $\chi^2$ values in Table
\ref{tab:chi2_interspectra_atlas_8tev_lj}, we find that the $\chi^2$ of each of
the individual spectra improves
when considered as part of the combined fit, an effect that can partially be attributed to the fact that the $t\bar{t}$ data now receive a relatively larger weight with respect to the other processes in the fit. This provides a clear motivation for publishing information on the
inter-spectra correlations, and we hope in the future these will also be
provided for some of the other data sets. 

Note in this context that while for the
ATLAS $t\bar{t}$ 13 TeV measurement in the $\ell + j$ channel we have the covariance matrix for each spectrum, including both
statistical and systematic uncertainties, and the inter-spectra statistical correlations, we don't have the correlations of systematics across spectra in the form of an off-diagonal covariance matrix. These elements can in principle be inferred by assuming that contributions from the same source are fully correlated, and that the asymmetric uncertainties are symmetrized using a particular prescription. Pending clarification on both these issues, we postpone an examination of the inter-spectra correlations in this data set for future work.

\begin{table}
  \centering
  \renewcommand{\arraystretch}{1.2}
  \begin{tabular}{l|l|l}
  $\chi^2/n_{\rm dat}$&Individual spectra&Combined\\
  \toprule  
  $m_{t\bar{t}}$ & 0.20 & 0.29\\
  $p_T^t$ & 0.37 & 0.41\\
  $y_t$ & 4.15& 3.55\\
  $y_{t\bar{t}}$ & 4.20& 3.73\\
  \midrule
  Total &$-$ & 1.84\\
  \end{tabular}
  \vspace{0.3cm}
  \caption{The $\chi^2/n_{\rm dat} $ corresponding to
  Fig.~\ref{fig:interspectra} for the ATLAS 8 TeV $\ell + j$
  $t\bar{t} $ differential spectra, shown for each distribution when fitted 
  individually and in the combined analysis where inter-spectra correlations are
  considered.}
  \label{tab:chi2_interspectra_atlas_8tev_lj}
\end{table}

\subsection{Impact of other small corrections}
\label{subsec:as_mt_n3lo}
With our baseline results established at NNLO with MHOU, and the demonstration that 
the current cross-sections are actually sufficient to determine the top mass to a precision of a few tenths of a GeV, we now move to study
the impact of various small cumulative corrections. First, we consider aN$^3$LO QCD
corrections, followed by the addition of mixed QCD$\otimes$QED corrections and
the photon PDF. On top of this, we then consider EW corrections to the $t\bar{t}$ matrix
element, the toponium correction, and finally
the impact of the FLAG determination of $\alpha_s$. 

\begin{figure}[htbp]
    \centering
    \includegraphics[width=0.49\linewidth]{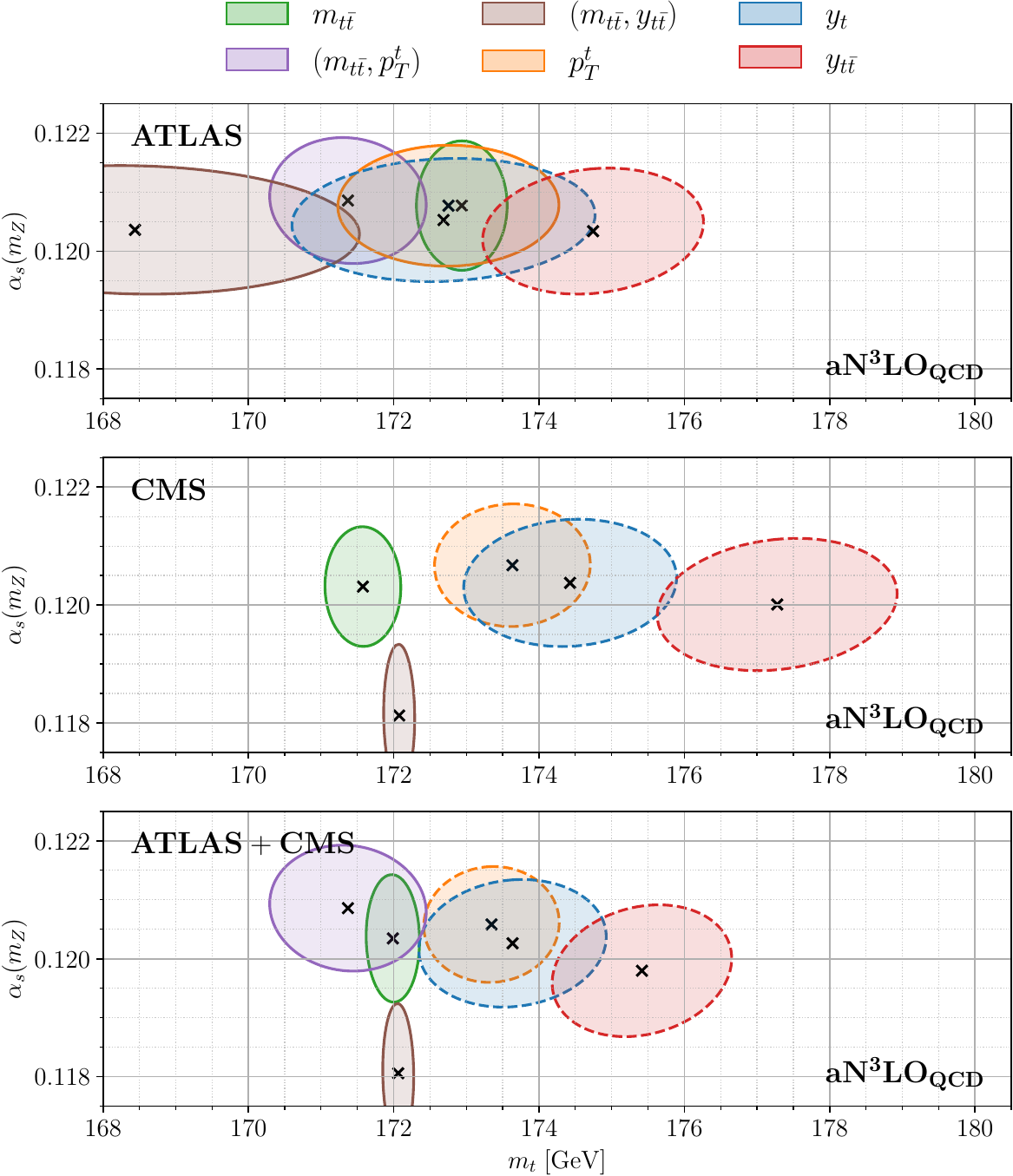}
    \includegraphics[width=0.49\linewidth]{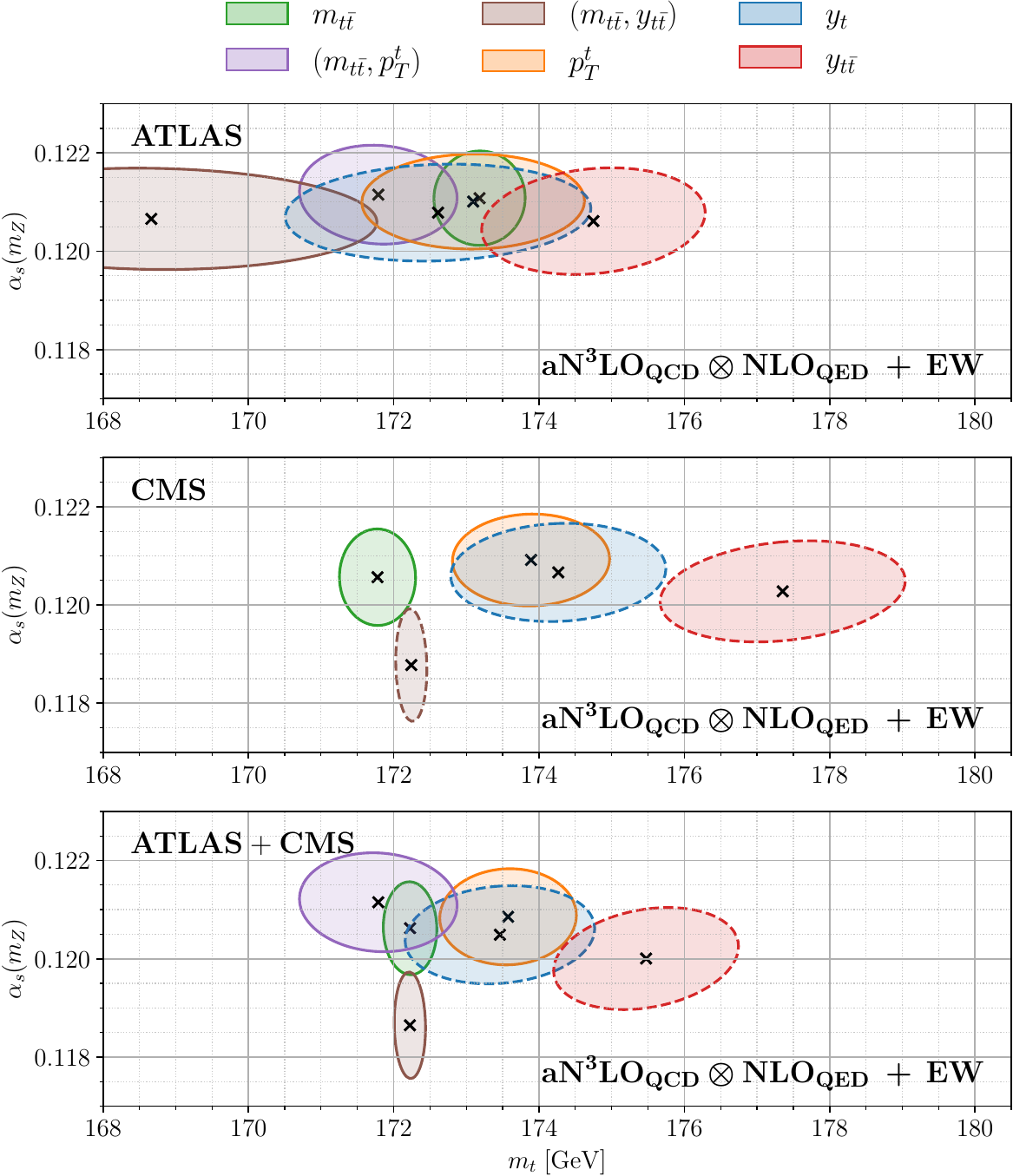}
    \caption{Left: same as Fig.~\ref{fig:obs-comparison-nnlo}
    \label{fig:obs-comparison-n3lo-qed}, now at ${\rm aN}^3{\rm LO}$ QCD
    accuracy. Right: same, now at ${\rm aN}^3{\rm LO}$ QCD accuracy including
    mixed QCD$\otimes$QED PDF evolution, as well as EW corrections at the level of the
    $t\bar{t}$ matrix element.}
\end{figure}

\paragraph{aN$^{3}$LO corrections} The left panel of
Fig.~\ref{fig:obs-comparison-n3lo-qed} shows the impact of a${\rm N}^3{\rm LO}$
QCD evolution effects in the $(\alpha_s, m_t)$ plane. The corresponding
numerical bounds on $m_t$ are collected in Table \ref{tab:mt_bounds_overview}. Comparing these
against the NNLO results in that same table, we observe
how a${\rm N}^3{\rm LO}$ perturbative corrections push the top mass down for
nearly all
observables, with $m_t$ moving down by around $0.3$ GeV in the case of
$m_{t\bar{t}}$, while rapidity based extractions push $m_t$ down almost by 1
GeV. This effect we have narrowed down to a decreased gluon luminosity at a${\rm
N}^3{\rm LO}$ around the $t\bar{t}$ threshold, which the fit compensates by
preferring a higher partonic cross-section, which in turn is realised at lower
values of the top mass. We do not observe any significant shift
in $\alpha_s$, which confirms that our MHOUs at NNLO QCD correctly account for higher
order effects. This was previously also observed in Ref.~\cite{Ball:2025xgq}. Regarding the fit quality, from Table
\ref{tab:chi2-no-toponium} we can conclude that the $\chi^2$ to the top datasets is little changed on going from NNLO to 
${\rm aN}^3{\rm LO}$ QCD, as we would expect given that in both cases the top cross-section is evaluated at NNLO with an estimate for MHOUs. 

Beyond NNLO QCD, one may also ask about the impact of soft gluon resummation effects on our
$m_t$ determination. In the case of the CMS dilepton measurement at 13 TeV from
Ref.~\cite{CMS:2018adi}, NNLL$^\prime$ corrections have been computed in
Ref.~\cite{Czakon:2019txp}, and we have verified explicitly that these lead to no noticeable change in our $m_t$
determination, confirming that our MHOUs already capture these higher
order effects. We therefore do not consider any soft gluon resummation effects
in this work. However, if NNLL$^\prime$ corrections become available for the other datasets, it could be interesting to
revisit their (combined) impact.  We note that threshold effects may become more important if experimental resolution around threshold improves in the future.

It would also be interesting in the future to attempt to construct
an approximation to the ${\rm N}^3{\rm LO}$ differential top cross-sections by combining the results from soft gluon resummation with those from high energy resummation~\cite{Collins:1991ty,Ball:2001pq,Silvetti:2022hyc},
along the lines of the ${\rm aN}^3{\rm LO}$ result already obtained for the total cross-section~\cite{Muselli:2015kba}. Such an approximation might  potentially increase the precision and accuracy of our top mass determination.

\begin{table}[htbp]
  \centering
  \footnotesize
  \renewcommand{\arraystretch}{1.4}
  \begin{tabular}{l|c|c|c}
&$m_t$ [GeV]&$m_t$ [GeV]&$m_t$ [GeV]\\
\toprule
Observable & $\bm{\mathrm{NNLO}_{\mathrm{QCD}}}$ 
& $\bm{\mathrm{aN^3LO}_{\mathrm{QCD}}}$
& \makecell{$\bm{\mathrm{aN^3LO}_{\mathrm{QCD}}\otimes
\mathrm{NLO}_{\mathrm{QED}}}$} \\
\midrule
{$m_{t\bar{t}}$} & $172.31 \pm 0.25$ & $171.99 \pm 0.24$ & $171.95 \pm 0.25$ \\
{$p_T^t$} & $174.04 \pm 0.63$ & $173.35 \pm 0.62$ & $173.25 \pm 0.62$ \\
{$y_t$} & $174.61 \pm 0.88$ & $173.64 \pm 0.85$ & $173.52 \pm 0.87$ \\
{$y_{t\bar{t}}$} & $176.47 \pm 0.84$ & $175.42 \pm 0.82$ & $175.34 \pm 0.83$ \\
{$(m_{t\bar{t}}, p_T^t)$} & $171.67 \pm 0.70$ & $171.37 \pm 0.71$ & $171.39 \pm 0.72$ \\
{$(m_{t\bar{t}}, y_{t\bar{t}})$} & $172.00 \pm 0.16$ & $172.06 \pm 0.14$ & $172.00 \pm 0.14$ \\
\midrule
& \makecell{$\bm{\mathrm{aN^3LO}_{\mathrm{QCD}}\otimes
\mathrm{NLO}_{\mathrm{QED}}}$\\+ \textbf{EW}}
& \makecell{$\bm{\mathrm{aN^3LO}_{\mathrm{QCD}}\otimes
\mathrm{NLO}_{\mathrm{QED}}}$\\ + \textbf{EW} + \textbf{topo.}}
& \makecell{$\bm{\mathrm{aN^3LO}_{\mathrm{QCD}}\otimes
\mathrm{NLO}_{\mathrm{QED}}}$\\ + \textbf{EW} + \textbf{topo.} + \textbf{FLAG}}\\
\midrule
{$m_{t\bar{t}}$} & $172.23 \pm 0.24$ & $172.81 \pm 0.26$ & $172.83 \pm 0.27$ \\
{$p_T^t$} & $173.57 \pm 0.62$ & $173.77 \pm 0.64$ & $173.76 \pm 0.66$ \\
{$y_t$} & $173.46 \pm 0.86$ & $173.61 \pm 0.86$ & $173.53 \pm 0.89$ \\
{$y_{t\bar{t}}$} & $175.47 \pm 0.84$ & $175.60 \pm 0.84$ & $175.61 \pm 0.85$ \\
{$(m_{t\bar{t}}, p_T^t)$} & $171.79 \pm 0.72$ & $172.01 \pm 0.72$ & $172.08 \pm 0.74$ \\
{$(m_{t\bar{t}}, y_{t\bar{t}})$} & $172.22 \pm 0.14$ & $172.96 \pm 0.24$ & $172.97 \pm 0.24$ \\
\midrule
\end{tabular}

  \vspace{0.3cm}
  \caption{Overview of the 68\% C.L. bounds on the top-quark pole mass, $m_t$, for
  each of the observables and theory settings considered in this work. The upper
  half consists of, first, our baseline NNLO QCD theory, followed by the addition of
  aN$^3$LO QCD corrections to the PDF evolution, and mixed
  ${\rm QCD}\otimes{\rm QED}$ evolution. The lower half then additionally
  includes, first, EW corrections to the $t\bar{t}$ matrix element, a toponium
  correction, and finally the FLAG constraint imposed as prior on
  $\alpha_s(m_Z)$.}
  \label{tab:mt_bounds_overview}
\end{table}

\paragraph{QED evolution and EW corrections} We now move to study the impact on
the top mass of,
first, joint QCD$\otimes$QED evolution and the
photon PDF, followed by the impact of additionally including EW corrections in the hard
$t\bar{t}$ matrix element. 

Considering first the impact of QED evolution corrections, we provide in the
top-right column of Table
\ref{tab:mt_bounds_overview} the corresponding
bounds on $m_t$.  With respect to the
aN$^3$LO result we observe only minor differences,
which is expected given that the $t\bar{t}$ hard matrix element in this
particular fit is evaluated at NNLO QCD accuracy
without any EW corrections in the $t\bar{t}$ hard matrix element. Let us
therefore analyse now explicitly the impact of EW corrections as discussed in
Sect.~\ref{subsec:theory_settings}. The corresponding bounds are provided in the
bottom-left column of Table \ref{tab:mt_bounds_overview}. Compared to the
equivalent setup without EW corrections (top-right column), we note how the EW
corrections push $m_t$ up by around 0.3 GeV in the case of $m_{t\bar{t}}$ and
$p_T^t$ based distributions, while little effect is seen in the case of rapidity
based distributions. This is consistent with the relative size of the EW
corrections. Indeed, recalling Fig.~\ref{fig:ewk-factors}, we observe how EW corrections modify the $m_{t\bar{t}}$ and
$p_T^t$ distributions at the percent level, while only per-mille
level effects are observed in
the case of rapidity distributions. 

In the leftmost results column labeled ``+EW'' in Table \ref{tab:chi2-topo-vs-notopo}, we present the $\chi^2$
values while including EW corrections for the ATLAS and CMS combination. As
compared to the same setup without EW corrections, see Table
\ref{tab:chi2-no-toponium}, we find that including EW corrections
improves the fit
quality of the $m_{t\bar{t}}$ and $p_T^t$
distributions, reducing the $\chi^2$ from
1.163 to 1.133 and from 1.069 to 1.053, respectively. A slight deterioration is
observed instead in the case of the rapidity distributions, which could possibly
be explained by an increased tension with the jet data~\cite{Ball:2025xgq}.

Regarding their combined impact on $\alpha_s(m_Z)$, after comparing the left and
right panels in Fig.~\ref{fig:obs-comparison-n3lo-qed} we find that QED evolution and EW corrections lead to an
increase of $\alpha_s(m_Z)$ by a few per-mille. This is similar to the effect that was
also observed in Ref.~\cite{Ball:2025xgq}, which was explained by noting how the
photon PDF subtracts momentum from the gluon PDF which in turn is compensated by
a slightly higher value of $\alpha_s$. 

\begin{table}[htbp]
  \centering
  \scriptsize
  \renewcommand{\arraystretch}{1.2}
  \begin{tabular}{llc|ccc}   &   &  &
\multicolumn{3}{c}{aN$^3$LO$_{\rm QCD}\otimes$NLO$_{\rm QED}$} \\
\cmidrule(lr){1-6} 
Observable & Data set & $ n_{\rm dat} $ & + EW & \makecell{+ EW\\+ toponium} & \makecell{+ EW\\+
toponium\\+ FLAG} \\ 
\toprule 
\multirow[t]{7}{*}{$m_{t\bar{t}}$} & ATLAS 13~TeV $t\bar{t}$ all hadr. & 9 & 1.049 & 1.051 & 1.039 \\
 & ATLAS 13~TeV $t\bar{t}$ $\ell+j$ & 9 & 0.798 & 0.837 & 0.778 \\
 & ATLAS 8~TeV $t\bar{t}$ $2\ell$ & 6 & 0.064 & 0.053 & 0.039 \\
 & ATLAS 8~TeV $t\bar{t}$ $\ell+j$  & 7 & 0.362 & 0.417 & 0.519 \\
 & CMS 13~TeV $t\bar{t}$ $2\ell$ 138 ${\rm fb}^{-1}$ & 7 & 1.419 & 1.268 & 1.296 \\
 & CMS 13~TeV $t\bar{t}$ $\ell+j$ & 15 & 0.879 & 0.852 & 0.985 \\
 & total & 53 & 1.133 & 1.055 & 1.055 \\
\cline{1-6}
\multirow[t]{2}{*}{$(m_{t\bar{t}}, p_T^t)$} & ATLAS 13~TeV $t\bar{t}$ $\ell+j$ & 15 & 0.490 & 0.530 & 0.548 \\
 & total & 15 & 0.490 & 0.530 & 0.548 \\
\cline{1-6}
\multirow[t]{4}{*}{$(m_{t\bar{t}}, y_{t\bar{t}})$} & ATLAS 13~TeV $t\bar{t}$ all hadr. & 11 & 1.177 & 1.173 & 1.239 \\
 & CMS 13~TeV $t\bar{t}$ $2\ell$ 138 ${\rm fb}^{-1}$ & 16 & 1.873 & 1.879 & 1.847 \\
 & CMS 13~TeV $t\bar{t}$ $\ell+j$ & 34 & 2.849 & 2.793 & 2.832 \\
 & total & 61 & 2.415 & 2.389 & 2.405 \\
\cline{1-6}
\multirow[t]{5}{*}{$p_T^t$} & ATLAS 13~TeV $t\bar{t}$ $\ell+j$ & 8 & 0.757 & 1.166 & 1.091 \\
 & ATLAS 8~TeV $t\bar{t}$ $\ell+j$  & 8 & 0.434 & 0.581 & 0.547 \\
 & CMS 13~TeV $t\bar{t}$ $2\ell$ 138 ${\rm fb}^{-1}$ & 7 & 2.924 & 2.694 & 2.617 \\
 & CMS 13~TeV $t\bar{t}$ $\ell+j$ & 16 & 0.497 & 0.655 & 0.623 \\
 & total & 39 & 1.053 & 1.155 & 1.117 \\
\cline{1-6}
\multirow[t]{5}{*}{$y_t$} & ATLAS 13~TeV $t\bar{t}$ $\ell+j$ & 5 & 0.725 & 0.709 & 0.777 \\
 & ATLAS 8~TeV $t\bar{t}$ $\ell+j$  & 5 & 4.331 & 4.152 & 4.188 \\
 & CMS 13~TeV $t\bar{t}$ $2\ell$ 138 ${\rm fb}^{-1}$ & 10 & 3.088 & 3.092 & 3.046 \\
 & CMS 13~TeV $t\bar{t}$ $\ell+j$ & 11 & 3.122 & 3.124 & 2.984 \\
 & total & 31 & 2.903 & 2.882 & 2.828 \\
\cline{1-6}
\multirow[t]{7}{*}{$y_{t\bar{t}}$} & ATLAS 13~TeV $t\bar{t}$ all hadr. & 12 & 0.727 & 0.730 & 0.727 \\
 & ATLAS 13~TeV $t\bar{t}$ $\ell+j$ & 7 & 0.434 & 0.443 & 0.431 \\
 & ATLAS 8~TeV $t\bar{t}$ $2\ell$ & 5 & 0.762 & 0.747 & 0.774 \\
 & ATLAS 8~TeV $t\bar{t}$ $\ell+j$  & 5 & 2.652 & 2.565 & 2.711 \\
 & CMS 13~TeV $t\bar{t}$ $2\ell$ & 10 & 0.792 & 0.771 & 0.838 \\
 & CMS 13~TeV $t\bar{t}$ $\ell+j$ & 10 & 3.985 & 3.954 & 3.843 \\
 & total & 49 & 1.686 & 1.662 & 1.669 \\
\bottomrule
\end{tabular}

  \vspace{0.3cm}
  \caption{The $\chi^2/n_{\rm dat}$ values corresponding to the combined fits to ATLAS and
  CMS data at aN$^3$LO$_{\rm QCD}\otimes{\rm NLO}_{\rm QED}$ supplemented successively by
  EW and toponium corrections, and finally the FLAG prior on $\alpha_s(m_Z)$. }
  \label{tab:chi2-topo-vs-notopo}
\end{table}

\paragraph{Toponium corrections}
\label{subsec:topo_results}

\begin{figure}[htbp]
    \centering
    \includegraphics[width=.49\linewidth]{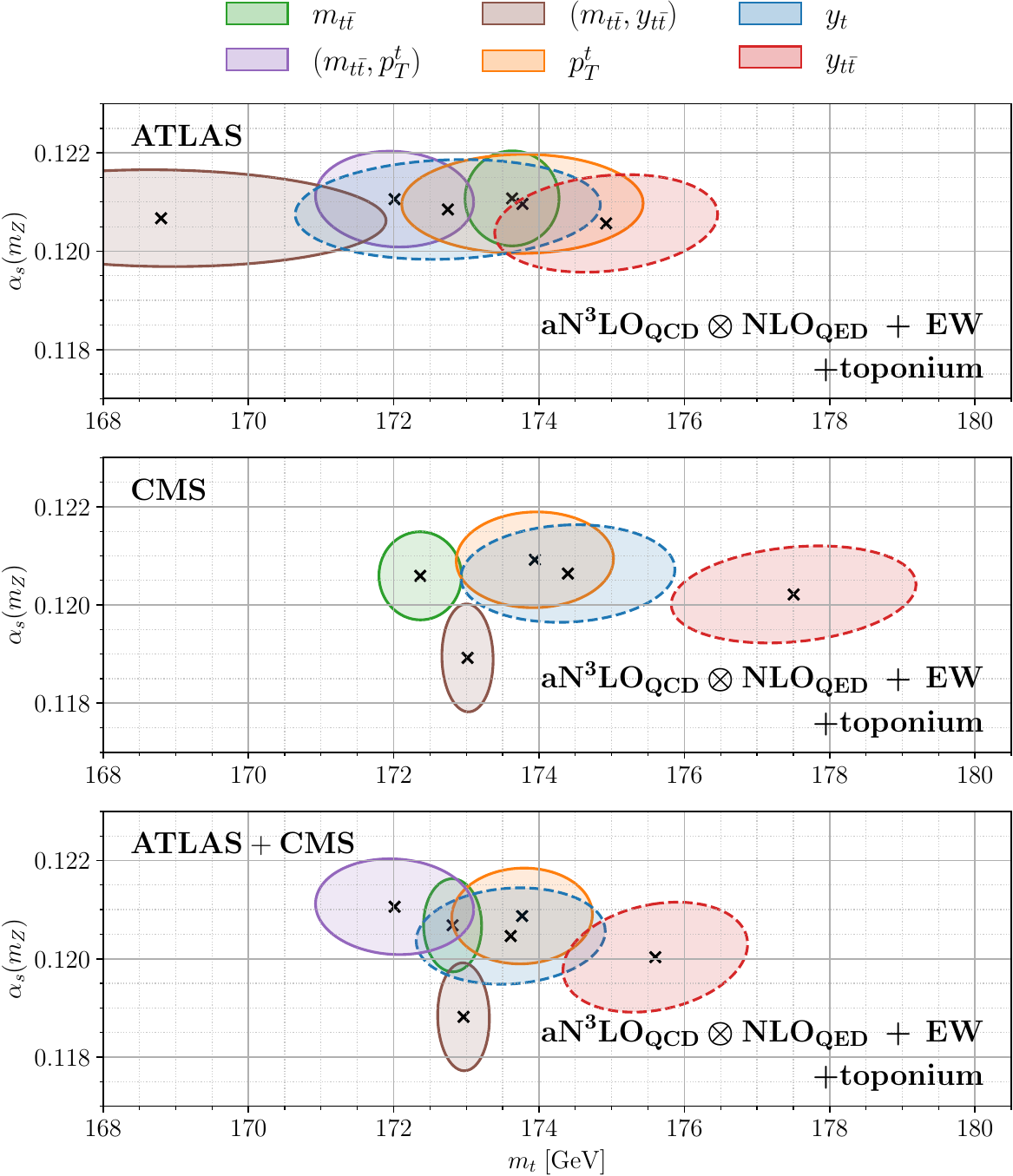}
    \includegraphics[width=.49\linewidth]{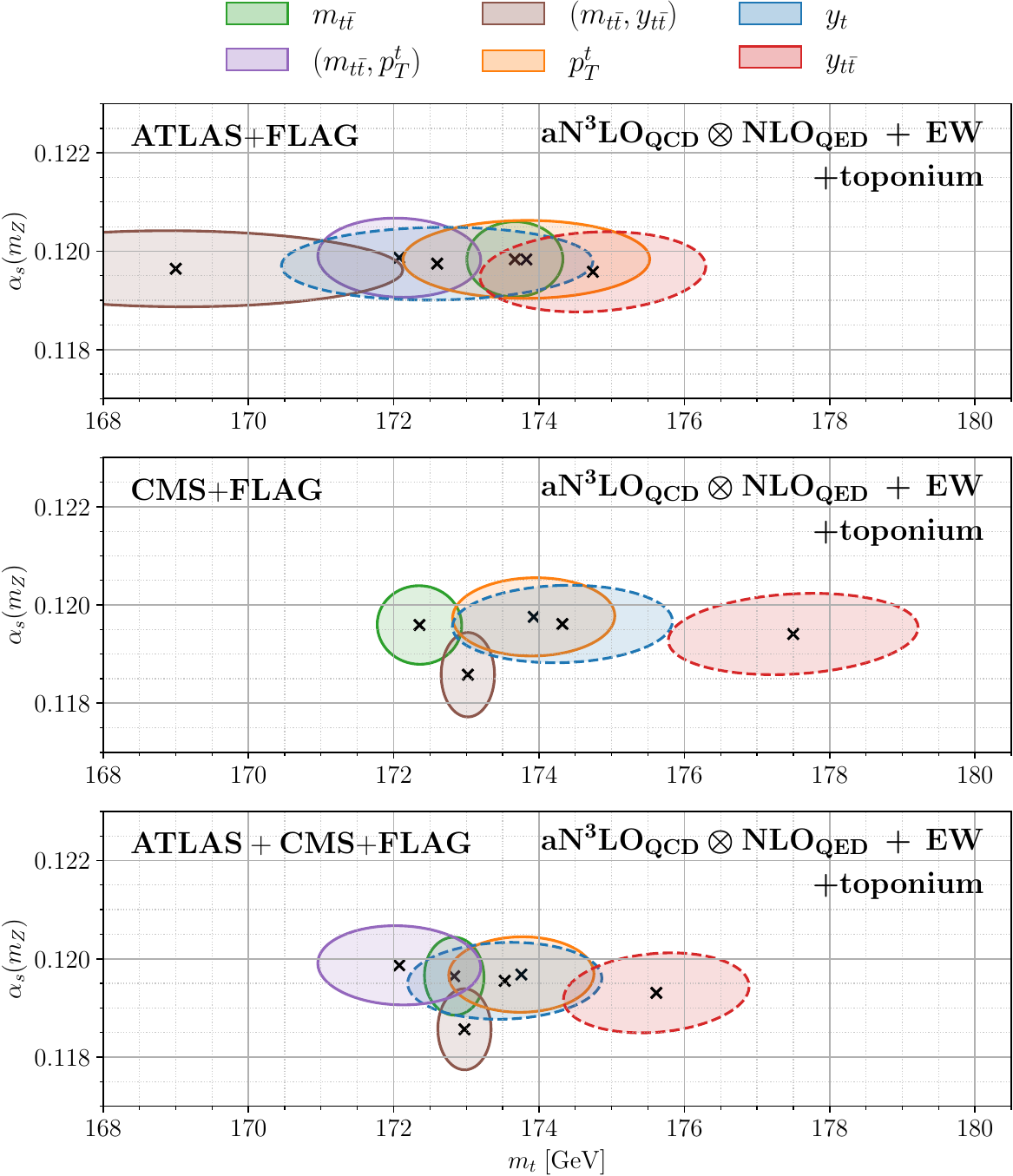}
    \caption{Left: same as the right panel of Fig.~\ref{fig:obs-comparison-n3lo-qed}, now in the
    presence of toponium corrections. Right: same, except we now set the
    $\alpha_s(m_Z)$ prior to the FLAG24 constraints
   ~\cite{FlavourLatticeAveragingGroupFLAG:2024oxs}.}
    \label{fig:obs-comparison-n3lo-qed-toponium}
\end{figure}

To assess quantitatively the impact of toponium corrections we display in the
left panel of Fig.~\ref{fig:obs-comparison-n3lo-qed-toponium} the 68\% C.L.
bounds in the presence of toponium corrections when added on top of the
aN$^3$LO$_{\rm QCD}\otimes{\rm NLO_{QED}}$ theory with EW
corrections. The numerical bounds are provided in Table
\ref{tab:mt_bounds_overview} (lower half, second column). As discussed in
Sect.~\ref{subsec:toponium_corrections}, this includes a 50\% theoretical
uncertainty on the total toponium cross-section. 

Compared to the
equivalent setup without toponium corrections (lower half, first column), we
observe that toponium generally moves $m_t$ up by about 0.6 GeV in
the case of the single differential distributions in $m_{t\bar{t}}$,
corresponding to about 1.7$\sigma$, while $m_t$ receives an upward shift of 0.7 GeV in the case of the the double differential distribution in
$(m_{t\bar{t}}, y_{t\bar{t}})$, corresponding to about
2.7$\sigma$. We also note how the uncertainty on $m_t$ generally increases when
toponium corrections are included, as expected, since the toponium
theory covariance matrix in Eq.~\eqref{eq:topo_covmat} gives $m_t$
greater freedom to move up and down in the bins around the $t\bar{t}$ threshold. By contrast, toponium corrections have no significant impact in determinations
based on the rapidity distributions $y_t$ and $y_{t\bar{t}}$. This is in line
with the fact that the toponium signal appears localised around threshold
in the $m_{t\bar{t}}$ distributions, while it contributes more uniformly across the
rapidity bins, thereby washing out its effect. In all cases, the uncertainties
do not change appreciably. As a result of this upward shift, the overall
compatibility across the different observables improves. With the exception of
the $y_{t\bar{t}}$ based
determination, all observables now lead to values of $m_t$ whose one sigma uncertainties all overlap. 

Regarding the fit quality, from inspecting Table
\ref{tab:chi2-topo-vs-notopo} we observe that the total $\chi^2$ of the $t\bar{t}$ datasets combined generally improves
under the addition of toponium with the exception of the $p_T^t$ based
determination, which suggest that overall the $t\bar{t}$ data sets prefer the
addition of a toponium correction.

\paragraph{FLAG determination of $\boldsymbol{\alpha_s}$}
\label{subsec:flag_results}
The recent lattice combination from FLAG
\cite{McNeile:2010ji, Chakraborty:2014aca, DallaBrida:2022eua, Petreczky:2020tky, Ayala:2020odx, Bazavov:2019qoo, Cali:2020hrj, Bruno:2017gxd, PACS-CS:2009zxm, Maltman:2008bx, FlavourLatticeAveragingGroupFLAG:2024oxs} provides a
stringent bound on $\alpha_s(m_Z)$, namely $\alpha_s(m_Z) = 0.1183 \pm 0.0007$. 
In light of the fact that this is considerably lower than most of our $\alpha_s$
determinations, it is interesting to
analyse to what extent this might affect our $m_t$ determination. 

Our starting point is the observation that the lattice result relies on input
data independent of any data included in our PDF fits, meaning that it can be safely
incorporated into our fit without double counting. To this end, we impose a
prior on $\alpha_s(m_Z)$ around $\alpha_s(m_Z) = 0.1183 \pm 0.0007$ and generate
theoretical predictions at $\alpha_s(m_Z)=\{0.1176, 0.1183, 0.1190\}$ in order
to construct the corresponding theory covariance matrix as discussed in
Sect.~\ref{subsec:tcm_method}. We then repeat our analysis at aN$^3$LO$_{\rm
QCD}\otimes{\rm NLO_{QED}}$ with EW corrections and toponium. The result is displayed in the right panel of
Fig.~\ref{fig:obs-comparison-n3lo-qed-toponium}. We can clearly see how $\alpha_s(m_Z)$
now comes out considerably lower, although the experimental data in the PDF fit still prefer a
value of $\alpha_s(m_Z)$ higher than FLAG by about $1.5\sigma$. Importantly,
inspecting Table \ref{tab:mt_bounds_overview}, we see no
significant shifts in our $m_t$ determinations, as expected given the relatively
weak correlation of $m_t$ with $\alpha_s$. The same holds for the $\chi^2$
values, as is clear from the rightmost column in Table \ref{tab:chi2-topo-vs-notopo}.

\subsection{Comparison to other determinations of the top mass}
\label{subsec:comp_lit}
Finally, we comment, whenever possible, on how our results compare with some
previous determinations. In particular, we compare with
cross-section based measurements and direct
reconstruction measurements from ATLAS and CMS, the ABMPtt determination~\cite{Alekhin:2024bhs},
the MSHT determination~\cite{Cridge:2023ztj}, and the analysis of
Ref.~\cite{Cooper-Sarkar:2020twv} that adopts a fixed PDF set. 

\begin{figure}[htbp]
  \centering
  \includegraphics[width=0.85\linewidth]{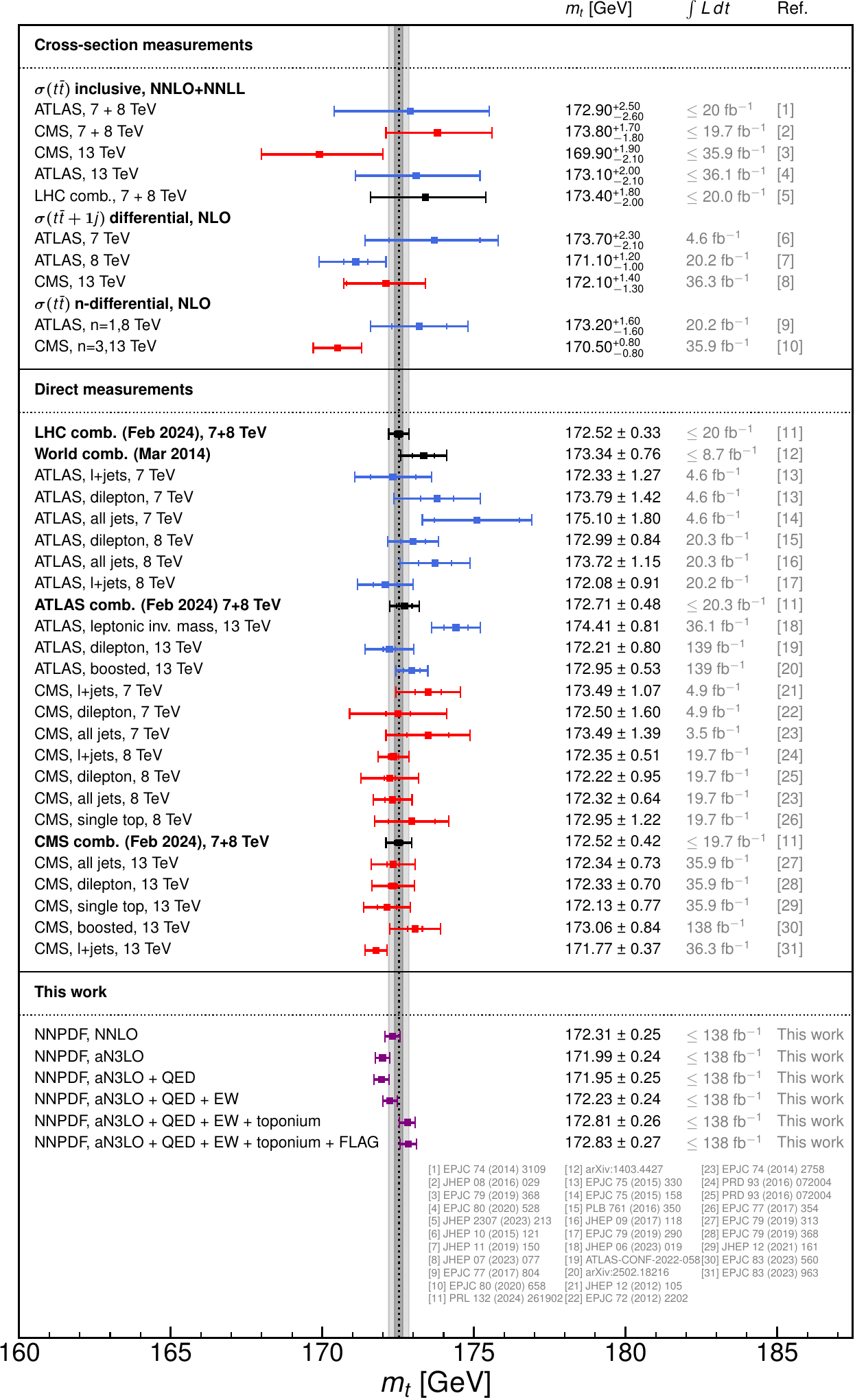}
  \caption{A comparison between cross-section based measurements and direct
  reconstruction measurements by ATLAS and CMS versus our determination based on
  cross-section measurements differential in the invariant mass of the top-quark
  pair. The grey band denotes the LHC combination from Feb 2024, which combines 7 and 8 TeV data from ATLAS and CMS. The plot is adapted from the LHCTopWG top mass summary plots, May 2025.}
  \label{fig:direct_reconstruction_comp}
\end{figure}

\paragraph{ATLAS and CMS} Fig.~\ref{fig:direct_reconstruction_comp} shows a
comparison of our determination of $m_t$ with those reported by ATLAS and CMS. Results are grouped into indirect measurements (upper
panel), which are based on similar  
measurements of $t\bar{t}$ production cross-sections to those used in our own determination, and direct measurements (middle panel), which
are based instead on reconstruction of the top mass from measurements of the kinematics of the final state after the top quarks decay. While the direct measurements are significantly more precise, they rely rather heavily on Monte Carlo event generators, and thus suffer from a potential ambiguity in the precise definition of the top mass.

In the lower panel, we indicate
the global determinations we obtained in the current work, using the various theoretical
variations that we considered. We have restricted this to invariant mass based
determinations given that these show the best compromise between precision and
goodness of fit. Inspecting Fig.~\ref{fig:direct_reconstruction_comp}, it is
remarkable that our indirect cross-section based measurement is very
consistent with the direct measurements, and even a little bit more precise.
This can be explained by noting that our fit combines multiple ATLAS and CMS
measurements from different data sets while
simultaneously taking account of correlations through the PDFs and $\alpha_s$. 

\paragraph{ABMPtt}
Next, we compare our results at NNLO QCD with MHOUs to the simultaneous ABMPtt
determination from Ref.~\cite{Alekhin:2024bhs} (see in particular Table I
therein). Our analysis differs in many respects from that used by ABMPtt:  they also fit the PDFs together with $m_t$ and $\alpha_s$, but using a fixed functional form; their treatment of heavy quarks is different, and they do not account for MHOUs. Moreover no jet data is included in the
ABMPtt determination, which is especially relevant given the fact that this together with the top data is largely what determines the gluon PDF~\cite{Ball:2025xgq,
Ball:2025xtj}. Finally Ref.~\cite{Alekhin:2024bhs} determines the $\overline{\rm
MS}$ mass $m_t^{\overline{\rm MS}}$ rather than the pole mass: we convert $m_t^{\overline{\rm MS}}$ to the pole mass at four loops in QCD
in order to make a numerical comparison (see
Ref.~\cite{ParticleDataGroup:2024cfk} and in particular Eq.~(61.1) therein).

We include in Table \ref{tab:mt_alphas_comparison} a comparison of the top data sets
that enter both analyses. For the purpose of this comparison, we have
rerun our analysis using normalised differential $t\bar{t}$ distributions
rather than absolute ones. Comparing first the
ATLAS 13 TeV fully hadronic double differential measurement in
$(m_{t\bar{t}},y_{t\bar{t}})$, we find a compatible value for $m_t$
within uncertainties, with a comparable size of the uncertainty itself. However using 
the CMS 13 TeV measurement in the $\ell + j$ channel we find a value of $m_t$
around 4$\sigma$ higher than the one from ABMPtt,
with an uncertainty that is more than five times smaller. The precise reason for this substantial discrepancy 
is unclear.  

\paragraph{MSHT}
Ref.~\cite{Cridge:2023ztj} presented a simultaneous determination of the
top-quark pole mass $m_t$, $\alpha_s(m_Z)$, and the PDFs within the global MSHT PDF framework. They also use a given PDF parametrization, and have no MHOUs, but they do include inclusive jet data to constrain the gluon.
Their analysis focusses on top-quark data in the $\ell + j$ channel at 8 TeV
from ATLAS and CMS, each differential in $m_{t\bar{t}}$, $p_T^t$ and the
rapidities $y_t$ and $y_{t\bar{t}}$. No other channels were considered. In the case of ATLAS, they include all
observables simultaneously while accounting for inter-spectra correlations
through a specific decorrelation model~\cite{Bailey:2020ooq}. In the case of
CMS, they fit one observable at a time, since no inter-spectra
correlations are available for this data set. The corresponding theoretical
predictions are obtained with NNLO QCD supplemented with EW corrections. In line with our analysis, they observe only a very weak correlation between $ m_t$ and
$\alpha_s(m_Z)$. After an initial simultaneous fit, they exploit this weak correlation to perform $m_t$ determinations at
a fixed value of $\alpha_s(m_Z)=0.118$. We show in Table
\ref{tab:mt_alphas_comparison} the bounds on $m_t$ obtained by MSHT from a fit
to the ATLAS 8 TeV data set in the $\ell + j$ channel single differential in
$p_T^t$, $m_{t\bar{t}}$, $y_{t\bar{t}}$ and $y_t$ while considering
inter-spectra correlations at $\alpha_s=0.118$. This is consistent with our
determination displayed in Fig.~\ref{fig:interspectra} within one standard deviation, and with a
comparable size of the uncertainty. The final result reported by MSHT,
including also CMS differential data, is $m_t = 173.0
\pm 0.6$ GeV, fully compatible with our determination of $m_t=172.81 \pm 0.26$ GeV.

\begin{table}[t!]
  \centering
  \scriptsize
  \renewcommand{\arraystretch}{1.3}
  \begin{tabular}{l|l|l}
& \textbf{Other determinations}& \textbf{This work}  \\
\toprule
Data set and observable & \textbf{ABMPtt}~\cite{Alekhin:2024bhs} & \\
\midrule
ATLAS 13 TeV $t\bar{t}$ all hadr. & $m_t^{\overline{\rm MS}} = 160.5 \pm 2.0$ GeV &  $m_t^{\rm pole}=166.49 \pm 1.91$ GeV\\
$1/\sigma d\sigma/d(m_{t\bar{t}}, y_{t\bar{t}})$&  $m_t^{\rm pole} = 169.98 \pm 2.1 $ GeV&  $\alpha_s=0.12082 \pm 0.00069$  \\[6pt]
CMS 13 TeV $t\bar{t}$ $\ell + j$  & $m_t^{\overline{\rm MS}} = 158.7 \pm 0.9$ GeV& $m_t^{\rm pole}=171.70 \pm 0.14$ GeV\\
$1/\sigma d\sigma/d(m_{t\bar{t}}, y_{t\bar{t}})$&  $m_t^{\rm pole} = 168.09 \pm 1.0 $ GeV&  $\alpha_s=0.11906 \pm 0.00064$ \\
\midrule
 & \textbf{MSHT}~\cite{Cridge:2023ztj}&  \\
\midrule
ATLAS 8 TeV $t\bar{t}$ $\ell + j$ &$m_t^{\rm pole}=172.4$ GeV $< m_t < 173.6$
GeV & $m_t^{\rm pole} = 173.74 \pm 0.54$ GeV\\
$d\sigma/dp_T^t, d\sigma/dm_{t\bar{t}}, d\sigma/dy_{t\bar{t}}, d\sigma/dy_t$  &
$\alpha_s=0.118 $ (fixed)   &$\alpha_s=0.12059 \pm 0.00071$ \\
\midrule
 & \textbf{Ref.~\cite{Cooper-Sarkar:2020twv}} &   \\
\midrule
ATLAS 8 TeV $t\bar{t}$ $\ell + j$ &$m_t^{\rm pole}=175.6^{+1.0}_{-1.0}$ GeV&$m_t^{\rm pole} = 173.30 \pm 1.08$ GeV\\
$d\sigma/dp_T^t$ & $\alpha_s=0.1210^{+0.0024}_{-0.0023} $   &$\alpha_s=0.12059 \pm 0.00078$ \\[6pt]
ATLAS 8 TeV $t\bar{t}$ $\ell + j$ &$m_t^{\rm pole}=173.5^{+0.5}_{-0.5}$ GeV&$m_t^{\rm pole}=173.98 \pm 0.68$ GeV\\
$d\sigma/dm_{t\bar{t}}$& $\alpha_s=0.1176^{+0.0022}_{-0.0022} $   &$\alpha_s=0.12097 \pm 0.00067$ \\
\bottomrule
\end{tabular}
  \vspace{0.3cm}
  \caption{A comparison between a subset of our results (last column) and
  results from other PDF fitting groups. In particular, we compare our results to the
  work from ABMPtt~\cite{Alekhin:2024bhs}, MSHT~\cite{Cridge:2023ztj}, and the fixed PDF determination
  from Ref.~\cite{Cooper-Sarkar:2020twv}. We have indicated the pole-mass and
  $\overline{{\rm MS}}$ mass in superscript, for clarity.}
  \label{tab:mt_alphas_comparison}
\end{table}

\paragraph{Fixed PDF determinations}
Finally, we can also compare our results against determinations using fixed PDFs~\cite{Cooper-Sarkar:2020twv}, shown in the last two rows of Table \ref{tab:mt_alphas_comparison}. These analyse the ATLAS data taken at 8 TeV in the $\ell + j$ channel, 
using NNPDF3.1~\cite{NNPDF:2017mvq} for the PDFs, but performing a combined fit to $m_t$ and $\alpha_s$. 
Note that this approach neglects correlations with the PDFs, so we do not expect to find agreement a priori.
Using the single differential $p_T^t$ observable 
they find a result for $m_t$ a little higher than our own, but roughly compatible within uncertainties. Using the observable differential in 
$m_{t\bar{t}}$, we find statistically compatible results for 
$m_t$, while $\alpha_s(m_Z)$ comes out significantly lower in
Ref.~\cite{Cooper-Sarkar:2020twv}. This should come as no surprise given that the
determination in Ref.~\cite{Cooper-Sarkar:2020twv} ignores the very significant correlations between 
$\alpha_s(m_Z)$ and the PDFs~\cite{Forte:2020pyp,Forte:2025pvf}.
Note however that Ref.~\cite{Cooper-Sarkar:2020twv} finds a strong
(positive) correlation between $\alpha_s$ and $m_t$ when using observables differential in the
rapidity, which is also the observable with the highest degree of correlation in
our determination.

\section{Summary and outlook}
\label{sec:summary}

In this work we have presented a determination of the top-quark mass
while jointly varying the strong coupling constant within a global PDF analysis. We employed the theory covariance method (TCM), validated in this context using closure tests. We have considered a wide range of single and double differential $t\bar{t}$ cross-section measurements from ATLAS and CMS, taken at 8 and 13 TeV, 
analysing their individual and combined impact on the joint $(\alpha_s,
m_t)$ parameter space. The analysis is performed using NNLO QCD, including MHOUs, 
complemented with EW corrections, and accounting for PDF evolution up to
aN$^3$LO$_{\rm QCD}$$\otimes$NLO$_{\rm QED}$. We have  analysed, for the first time, the possible impact of toponium
corrections on the extraction of the top-quark mass. 

Our findings are as follows.
First, out of all the kinematic observables that we have considered, we find
that double differential measurements in $(m_{t\bar{t}}, y_{t\bar{t}})$ provide
the most stringent bounds on $m_t$, followed by single differential
distributions in $m_{t\bar{t}}$ and $p_T^t$. Observables differential in the top quark rapidity are the least
sensitive to $m_t$. In all cases, limited correlation between $\alpha_s$ and $m_t$ is observed, with the exception of rapidity based
determinations. The central value of $m_t$ is determined rather consistently across all the kinematic observables considered, generally within a standard deviation, and the outliers are observables which do not determine $m_t$ very precisely, and also have a relatively poor 
$\chi^2$ in the fit.  

Our most precise extraction with satisfactory
agreement between theory and data originates from the single
differential measurements in $m_{t\bar{t}}$: combining all the data from  ATLAS and CMS, we find $m_t = 172.31 \pm 0.25$ GeV at
NNLO with MHOUs, and $m_t = 172.81 \pm 0.26$ GeV at
aN$^3$LO$\otimes$NLO$_{\rm QED}$ with EW and toponium corrections, both in agreement with the current 
PDG average~\cite{ParticleDataGroup:2024cfk} and the LHC combination measurement
\cite{ATLAS:2024dxp}. It is particularly striking that our indirect measurement, by combining a variety of datasets in a fully correlated analysis, is both competitive with, and consistent with, the direct measurements from the kinematics of the final state, while not suffering from the well known Monte Carlo ambiguities inherent in such determinations.

Regarding the impact of perturbative corrections, we find that aN$^3$LO
corrections cause a slight downward shift of
$m_t$, while EW corrections push $m_t$ up
again by around 0.3 GeV. The overall picture that
emerges is that the agreement between results from different observables improves under the
addition of these higher order corrections. Toponium corrections have a rather greater impact, increasing $m_t$ by as much as 0.6 GeV
in extractions based on $m_{t\bar{t}}$. However 
$m_t$ instead remains relatively stable in
determinations differential in $p_T^t$ or the rapidities $y_t$ and
$y_{t\bar{t}}$. As a result, the toponium corrections
also improve the mutual consistency across kinematic observables. 

Finally, we have explicitly demonstrated how combining individual measurements for the purpose of determining SM quantities requires one to refit the PDFs in
order to correctly account for all correlations and avoid double counting. As is well known, a naive statistical a posteriori average underestimates uncertainties, as it ignores correlations,
while a PDF4LHC-like combination at the level of replicas significantly overestimates uncertainties, as does an envelope approach. The TCM takes all correlations into account correctly, making it the ideal method for performing such combinations, which is especially relevant when combining existing LHC determinations.  Here we also
note the benefit one gains once inter-spectra correlations across different
observables and hope that these will become more widely available in the future.

The techniques used here can be extended in various  directions. It will be
interesting to study how particle level, rather than parton level, top production measurements
affect the sensitivity to the top mass~\cite{ATLAS:2023gsl, ATLAS:2025iza}, by determining power corrections to the top cross-section using similar techniques to those recently explored for jets~\cite{Ball:2025xtj}.

The TCM methodology can be easily used to perform fully correlated global  determinations of additional SM
parameters, in particular the $W$-boson mass and the weak mixing angle $\sin^2\theta_W$.
As is well known, the largest sources of uncertainty in determinations of both these quantities at the LHC are the strong coupling and the PDFs. Moreover the $W$-boson mass receives radiative corrections sensitive to the top quark's mass, which can therefore introduce a non-trivial interplay~\cite{deBlas:2022hdk}. A truly precise global determination will require all these correlations to be properly accounted for. 

Beyond this, it will be interesting to attempt to fit, in addition, Wilson coefficients within the Standard Model Effective Field Theory (SMEFT)
framework~\cite{Cole:2026eex}. This is technically rather straightforward using the TCM, since the dependence of cross-sections on the SMEFT coefficients is easily computed, and any number of correlated coefficients can be determined within a single global PDF fit. Again the key to successful global determinations is to fully account for correlations, both to PDFs and to SM parameters. 
\bigskip 
\paragraph{Acknowledgments.}
We would like to thank the members of the NNPDF collaboration for insightful
discussions during the course of this work. We are particularly   grateful to Tanishq Sharma
for his work on the data set implementation and assistance with MATRIX in the
early stages of this project, and to Emanuele Nocera for assisting us with the electroweak corrections. J.t.H would like to thank Rafael Aoude and Thomas Cridge
for useful discussions, and we would also like to thank Paolo Nason and Luca Rottoli for correspondence on the estimation of the toponium correction. The work of R.D.B, J.t.H and R.S. is supported by
the Science and Technology Facilities Council (STFC) via grant awards
ST/T000600/1 and ST/X000494/1.
\bigskip
\appendix
\section{Toponium k-factors}
\label{sec:app-sup-res}
In this appendix, we present the numerical values of the toponium $k$-factors
that were computed following the procedure outline in
Sect.~\ref{subsec:toponium_corrections} for the data sets listed in Table
\ref{tab:datasets}. Table \ref{tab:kfactors-atlas} shows all the $k$-factors
different from unity in the case of ATLAS measurements, while Table
\ref{tab:kfactors-cms} shows the equivalent values for the CMS measurements. 

\begin{table}[t]
\centering
\begin{minipage}[t]{0.48\textwidth}
\centering
\small
\caption*{ATLAS $t\bar{t}$ 13 TeV hadr.}
\begin{tabular}{lll}
\toprule
Obs. & bin & $k_i^{\rm (topo)}$ \\
\midrule
$(m_{t\bar t},y_{t\bar t})$ & $((0,700),(0.00,0.46))$ & 1.0083 \\
                           & $((0,700),(0.46,0.91))$ & 1.0085 \\
                           & $((0,700),(0.91,1.55))$ & 1.0089 \\
                           & $((0,700),(1.55,2.50))$ & 1.0101 \\
\addlinespace
$y_{t\bar t}$              & $(0.00,0.12)$           & 1.0073 \\
                           & $(0.12,0.24)$           & 1.0072 \\
                           & $(0.24,0.36)$           & 1.0071 \\
                           & $(0.36,0.49)$           & 1.0072 \\
                           & $(0.49,0.62)$           & 1.0073 \\
                           & $(0.62,0.76)$           & 1.0073 \\
                           & $(0.76,0.91)$           & 1.0076 \\
                           & $(0.91,1.06)$           & 1.0074 \\
                           & $(1.06,1.21)$           & 1.0078 \\
                           & $(1.21,1.39)$           & 1.0079 \\
                           & $(1.39,1.59)$           & 1.0085 \\
                           & $(1.59,2.40)$           & 1.0098 \\
\addlinespace
$m_{t\bar t}$              & $(325.0,458.75)$        & 1.0167 \\
\bottomrule
\end{tabular}
\vspace{0.3cm}
\caption*{ATLAS $t\bar{t}$ 8 TeV $2\ell$.}
\begin{tabular}{lll}
\toprule
Obs & bin & $k_i^{\rm (topo)}$ \\
\midrule
$y_{t\bar t}$        & $(0.0,0.4)$ & 1.0082 \\
                     & $(0.4,0.8)$ & 1.0084 \\
                     & $(0.8,1.2)$ & 1.0089 \\
                     & $(1.2,2.0)$ & 1.0099 \\
                     & $(2.0,2.8)$ & 1.0133 \\
\addlinespace
$m_{t\bar t}$        & $(250.0,450.0)$ & 1.0184 \\
\bottomrule
\end{tabular}
\end{minipage}
\hfill
\begin{minipage}[t]{0.48\textwidth}
\centering
\small
\caption*{ATLAS $t\bar{t}$ 13 TeV $\ell + j$}
\begin{tabular}{lll}
\toprule
Obs. & bin & $k_i^{\rm (topo)}$ \\
\midrule
$m_{t\bar t}$        & $(325.0,400.0)$ & 1.0351 \\
\addlinespace
$p_T^t$              & $(0.0,50.0)$    & 1.0475 \\
\addlinespace
$y_t$                & $(0.0,0.4)$     & 1.0084 \\
                     & $(0.4,0.8)$     & 1.0083 \\
                     & $(0.8,1.2)$     & 1.0080 \\
                     & $(1.2,1.7)$     & 1.0076 \\
                     & $(1.7,2.5)$     & 1.0068 \\
\addlinespace
$y_{t\bar t}$        & $(0.0,0.25)$    & 1.0072 \\
                     & $(0.25,0.5)$    & 1.0072 \\
                     & $(0.5,0.8)$     & 1.0073 \\
                     & $(0.8,1.1)$     & 1.0075 \\
                     & $(1.1,1.4)$     & 1.0079 \\
                     & $(1.4,1.8)$     & 1.0087 \\
                     & $(1.8,2.5)$     & 1.0104 \\
\addlinespace
$(p_T,y_t)$          & $((0.0,0.75),(0.0,85.0))$ & 1.0229 \\
                     & $((0.75,1.5),(0.0,85.0))$ & 1.0213 \\
                     & $((1.5,2.5),(0.0,85.0))$  & 1.0171 \\
\addlinespace
$(m_{t\bar t},p_T^t)$ & $((325.0,500.0),(0.0,90.0))$ & 1.0237 \\
\bottomrule
\end{tabular}
\vspace{0.3cm}
\caption*{ATLAS $t\bar{t}$ 8 TeV $\ell + j$.}
\begin{tabular}{lll}
\toprule
Obs. & bin & $k_i^{\rm (topo)}$ \\
\midrule
$m_{t\bar t}$        & $(345.0,400.0)$ & 1.0207 \\
\addlinespace
$p_T^t$              & $(0.0,60.0)$    & 1.0364 \\
\addlinespace
$y_t$                & $(0.0,0.4)$     & 1.0099 \\
                     & $(0.4,0.8)$     & 1.0096 \\
                     & $(0.8,1.2)$     & 1.0091 \\
                     & $(1.2,1.6)$     & 1.0080 \\
                     & $(1.6,2.5)$     & 1.0061 \\
\addlinespace
$y_{t\bar t}$        & $(0.0,0.3)$     & 1.0082 \\
                     & $(0.3,0.6)$     & 1.0083 \\
                     & $(0.6,0.9)$     & 1.0085 \\
                     & $(0.9,1.3)$     & 1.0091 \\
                     & $(1.3,2.5)$     & 1.0105 \\
\bottomrule
\end{tabular}
\end{minipage}
\caption{Toponium k-factors for the ATLAS $t\bar t$ measurements listed in Table
\ref{tab:datasets}.}
\label{tab:kfactors-atlas}
\end{table}


\begin{table}[t]
\centering
\begin{minipage}[t]{0.48\textwidth}
\centering
\small
\caption*{(a) CMS $t\bar t$ 13 TeV $2\ell$, $138~\mathrm{fb}^{-1}$}
\begin{tabular}{lll}
\toprule
Obs. & bin & $k_i^{\rm (topo)}$ \\
\midrule
$m_{t\bar t}$        & $(300.0,380.0)$ & 1.0600 \\
\addlinespace
$p_T^t$              & $(0.0,55.0)$ & 1.0403 \\
\addlinespace
$y_t$                & $(-2.6,-1.8)$   & 1.0063 \\
                     & $(-1.8,-1.35)$  & 1.0075 \\
                     & $(-1.35,-0.9)$  & 1.0081 \\
                     & $(-0.9,-0.45)$  & 1.0081 \\
                     & $(-0.45,0.0)$   & 1.0084 \\
                     & $(0.0,0.45)$    & 1.0084 \\
                     & $(0.45,0.9)$    & 1.0082 \\
                     & $(0.9,1.35)$    & 1.0079 \\
                     & $(1.35,1.8)$    & 1.0075 \\
                     & $(1.8,2.6)$     & 1.0065 \\
\addlinespace
$(m_{t\bar t},y_{t\bar t})$
                     & $((300,400),(0.0,0.35))$ & 1.0365 \\
                     & $((300,400),(0.35,0.75))$ & 1.0362 \\
                     & $((300,400),(0.75,1.15))$ & 1.0358 \\
                     & $((300,400),(1.15,2.5))$  & 1.0346 \\
\bottomrule
\end{tabular}
\vspace{0.3cm}
\caption*{(c) CMS $t\bar t$ 13 TeV $\ell + j$}
\begin{tabular}{lll}
\toprule
Obs. & bin & $k_i^{\rm (topo)}$ \\
\midrule
$y_t$              & $(0.0,0.2)$ & 1.0084 \\
                   & $(0.2,0.4)$ & 1.0084 \\
                   & $(0.4,0.6)$ & 1.0083 \\
                   & $(0.6,0.8)$ & 1.0084 \\
                   & $(0.8,1.0)$ & 1.0080 \\
                   & $(1.0,1.2)$ & 1.0080 \\
                   & $(1.2,1.4)$ & 1.0077 \\
                   & $(1.4,1.6)$ & 1.0076 \\
                   & $(1.6,1.8)$ & 1.0075 \\
                   & $(1.8,2.0)$ & 1.0073 \\
                   & $(2.0,2.5)$ & 1.0063 \\
\addlinespace
$m_{t\bar t}$      & $(250.0,400.0)$ & 1.0350 \\
\addlinespace
$y_{t\bar t}$      & $(0.0,0.2)$ & 1.0073 \\
                   & $(0.2,0.4)$ & 1.0070 \\
                   & $(0.4,0.6)$ & 1.0074 \\
                   & $(0.6,0.8)$ & 1.0073 \\
                   & $(0.8,1.0)$ & 1.0076 \\
                   & $(1.0,1.2)$ & 1.0076 \\
                   & $(1.2,1.4)$ & 1.0080 \\
                   & $(1.4,1.6)$ & 1.0085 \\
                   & $(1.6,1.8)$ & 1.0090 \\
                   & $(1.8,2.4)$ & 1.0104 \\
\bottomrule
\end{tabular}
\end{minipage}
\hfill
\begin{minipage}[t]{0.48\textwidth}
\centering
\small
\caption*{(b) CMS $t\bar t$ 13 TeV $2\ell$ }
\begin{tabular}{lll}
\toprule
Obs. & bin & $k_i^{\rm (topo)}$ \\
\midrule
$p_T^t$              & $(0.0,65.0)$    & 1.0305 \\
\addlinespace
$m_{t\bar t}$        & $(300.0,380.0)$ & 1.0600 \\
\addlinespace
$y_t$                & $(-2.6,-1.8)$   & 1.0063 \\
                     & $(-1.8,-1.35)$  & 1.0075 \\
                     & $(-1.35,-0.9)$  & 1.0081 \\
                     & $(-0.9,-0.45)$  & 1.0081 \\
                     & $(-0.45,0.0)$   & 1.0084 \\
                     & $(0.0,0.45)$    & 1.0084 \\
                     & $(0.45,0.9)$    & 1.0082 \\
                     & $(0.9,1.35)$    & 1.0079 \\
                     & $(1.35,1.8)$    & 1.0075 \\
                     & $(1.8,2.6)$     & 1.0065 \\
\addlinespace
$y_{t\bar t}$        & $(-2.6,-1.6)$   & 1.0096 \\
                     & $(-1.6,-1.2)$   & 1.0083 \\
                     & $(-1.2,-0.8)$   & 1.0076 \\
                     & $(-0.8,-0.4)$   & 1.0072 \\
                     & $(-0.4,0.0)$    & 1.0071 \\
                     & $(0.0,0.4)$     & 1.0071 \\
                     & $(0.4,0.8)$     & 1.0073 \\
                     & $(0.8,1.2)$     & 1.0075 \\
                     & $(1.2,1.6)$     & 1.0081 \\
                     & $(1.6,2.6)$     & 1.0099 \\
\bottomrule
\end{tabular}
\vspace{0.3cm}
\caption*{(d) CMS $t\bar t$ 13 TeV $\ell + j$ (continued)}
\begin{tabular}{lll}
\toprule
Obs. & bin & $k_i^{\rm (topo)}$ \\
\midrule
$(m_{t\bar t},y_{t\bar t})$
                   & $((250.0,420.0),(0.0,0.3))$ & 1.0258 \\
                   & $((250.0,420.0),(0.3,0.6))$ & 1.0260 \\
                   & $((250.0,420.0),(0.6,0.9))$ & 1.0253 \\
                   & $((250.0,420.0),(0.9,1.2))$ & 1.0252 \\
                   & $((250.0,420.0),(1.2,1.5))$ & 1.0252 \\
                   & $((250.0,420.0),(1.5,2.5))$ & 1.0246 \\
\addlinespace
$p_T^t$             & $(0.0,40.0)$  & 1.0686 \\
                   & $(40.0,80.0)$ & 1.0010 \\
\bottomrule
\end{tabular}
\end{minipage}
\caption{Toponium k-factors for the CMS $t\bar t$ measurements listed in Table
\ref{tab:datasets}.}
\label{tab:kfactors-cms}
\end{table}


\clearpage

\bibliographystyle{JHEP}
\bibliography{alphas_mtop}

\end{document}